\documentclass[a4paper,11pt]{article} 
\pdfoutput=1 
\usepackage{jheppub}

\usepackage{amsmath, amssymb}
\usepackage{mathrsfs}
\usepackage{graphicx}
\usepackage{dcolumn}
\usepackage{bm}

\usepackage{comment}
\usepackage{braket}
\usepackage{appendix}
\usepackage{color,soul}
\usepackage{amsthm}
\usepackage{bbm}
\usepackage{bbm,bm,graphicx,mathtools,color,hyperref,slashed}
\usepackage[caption=false]{subfig}

\newcommand{\diff}{\mathrm{d}}

\newcommand{\Diff}{{\mathcal{D}}}

\newcommand{\tr}{\mathrm{tr}}
\newcommand{\im}{\mathrm{i}}

\newcommand{\rmd}{\mathrm{d}}
\newcommand{\rme}{\mathrm{e}}

\newcommand{\be}{\begin{equation}}      
\newcommand{\ee}{\end{equation}}      
\newcommand{\bea}{\begin{eqnarray}}      
\newcommand{\eea}{\end{eqnarray}}

\title{Semiclassical analysis of the bifundamental QCD on $\mathbb{R}^2\times T^2$ with 't~Hooft flux}

\author[]{Yui Hayashi,}
\emailAdd{yui.hayashi@yukawa.kyoto-u.ac.jp}

\author[]{Yuya Tanizaki,}
\emailAdd{yuya.tanizaki@yukawa.kyoto-u.ac.jp}

\author[]{and Hiromasa Watanabe}
\emailAdd{hiromasa.watanabe@yukawa.kyoto-u.ac.jp}

\affiliation[]{Yukawa Institute for Theoretical Physics, Kyoto University, Kitashirakawa Oiwakecho, Sakyo-ku, Kyoto 606-8502, Japan
}

\abstract{
We study the phase structure of bifundamental quantum chromodynamics (QCD(BF)), which is the $4$-dimensional $SU(N) \times SU(N)$ gauge theory coupled with the bifundamental fermion. 
Firstly, we refine constraints on its phase diagram from 't~Hooft anomalies and global inconsistencies, and we find more severe constraints than those in previous literature about QCD(BF).
Secondly, we employ the recently-proposed semiclassical approach for confining vacua to investigate this model concretely, and this is made possible via anomaly-preserving $T^2$ compactification. 
For sufficiently small $T^2$ with the 't~Hooft flux, the dilute gas approximation of center vortices gives reliable semiclassical computations, and we determine the phase diagram as a function of the fermion mass $m$, two strong scales $\Lambda_{1},\Lambda_2$, and two vacuum angles, $\theta_1, \theta_2$. 
In particular, we find that the QCD(BF) vacuum respects the $\mathbb{Z}_2$ exchange symmetry of two gauge groups. 
Under the assumption of the adiabatic continuity, our result successfully explains one of the conjectured phase diagrams in the previous literature and also gives positive support for the nonperturbative validity of the large-$N$ orbifold equivalence between QCD(BF) and $\mathcal{N}=1$ $SU(2N)$ supersymmetric Yang-Mills theory. 
We also comment on problems of domain walls. 
}

\preprint{YITP-23-96}

\begin{document}

\maketitle

\section{Introduction}

In this paper, we study the phase structure of Quantum Chromodynamics (QCD) with the bifundamental fermion (QCD(BF)). 
QCD(BF) is the $4$-dimensional $SU(N)_1 \times SU(N)_2$ gauge theory with a Dirac fermion in the bifundamental representation, and this theory may be regarded as the flavor-symmetric fundamental QCD with the same color and flavor numbers, $N_c=N_f=:N$, where the flavor symmetry is promoted to the gauge redundancy with the dynamical gauge fields. 
The flavor gauge fields can be turned off in the weak-coupling limit of the flavor gauge couplings, $g_2\to 0$, and the theory goes back to the usual QCD with the exact flavor symmetry. 

The parameters of QCD(BF) are the two strong scales, $\Lambda_1=\mu \rme^{-8\pi^2/(3N g_1^2(\mu))}$ and $\Lambda_2=\mu \rme^{-8\pi^2/(3N g_2^2(\mu))}$, two vacuum angles, $\theta_1$ and $\theta_2$, and the fermion mass $m$, 
and the phases of the QCD(BF) are expected to have rich structures as we change these parameters~\cite{Tanizaki:2017bam, Karasik:2019bxn}. 
The phase diagram has been investigated from the recent developments of the anomaly matching~\cite{tHooft:1979rat, Wen:2013oza, Kapustin:2014lwa, Kapustin:2014zva, Cho:2014jfa, Gaiotto:2017yup} or global inconsistency~\cite{Gaiotto:2017yup, Kikuchi:2017pcp, Tanizaki:2017bam, Tanizaki:2018xto, Karasik:2019bxn, Cordova:2019jnf, Cordova:2019uob} and also from various calculable limits. 
These studies are helpful in developing our knowledge about confinement and its relation to chiral symmetry breaking/the multi-branch $\theta$-angle structures, and they led us to the plausible proposal for the QCD(BF) phase diagram.

The other motivation behind the exploration of QCD(BF) comes from the concept of large-$N$ orbifold equivalence~\cite{Kachru:1998ys, Bershadsky:1998cb, Schmaltz:1998bg, Strassler:2001fs, Dijkgraaf:2002wr, Kovtun:2003hr, Kovtun:2004bz, Armoni:2005wta, Kovtun:2005kh}. 
The perturbative expansion in the large-$N$ limit is governed by the planar diagrams~\cite{tHooft:1973alw}, and, surprisingly, the $\mathcal{N}=1$ $SU(2N)$ supersymmetric Yang-Mills (SYM) theory and $SU(N)\times SU(N)$ QCD(BF) shares the same perturbative expansion at all orders of the 't~Hooft coupling when $N\to \infty$~\cite{Kachru:1998ys, Bershadsky:1998cb}.  
Interestingly, even though the QCD(BF) is non-supersymmetric, certain structures reminiscent of supersymmetry are anticipated due to this equivalence, and it is important to understand if the large-$N$ orbifold equivalence is true also for the nonperturbative physics. 
It has been proven that the nonperturbative equivalence holds if and only if both the orbifold-projection symmetry in the parent theory and the exchange symmetry of the $SU(N)\times SU(N)$ gauge group in the daughter theory remain unbroken~\cite{Kovtun:2003hr, Kovtun:2004bz}, and thus the correct identification of the ground states is also significant to settle the nonperturbative validity of the large-$N$ orbifold equivalence.

With these motivations, this paper studies QCD(BF) using the recently-proposed semiclassical description of the confining vacuum~\cite{Tanizaki:2022ngt}. 
In this semiclassical approach, we put the four-dimensional gauge theories on small $\mathbb{R}^2 \times T^2$ with 't Hooft flux, and it has been observed to successfully explain qualitative aspects of vacuum structures of various confining gauge theories~\cite{Tanizaki:2022ngt, Tanizaki:2022plm}, including $SU(N)$ Yang-Mills theory, $\mathcal{N} = 1$ SYM theory, QCD with fundamental quarks, and QCD with $2$-index quarks.
This approach stands out for its calculability and compatibility with widespread interpretations of quark confinement, particularly the center vortex picture~\cite{DelDebbio:1996lih, Faber:1997rp, Langfeld:1998cz, Kovacs:1998xm, Greensite:2011zz}.
In this paper, we apply this semiclassical computational method to find the phase diagram of QCD(BF) and discuss its relation with previous studies. 

The punchline of our results is the following. 
Within the validity of the semiclassics ($N\Lambda_{1,2}L\ll 1$ with the torus size $L$), we obtain the phase diagram not only as a function of $\theta_1$ and $\theta_2$ but also as the function of the fermion mass $m$ and the strong scales $\Lambda_{1,2}$, and our result covers from the chiral limit to the relatively heavy fermion mass.  
Moreover, it turns out that the $\mathbb{Z}_2$ exchange symmetry is unbroken, and our result provides affirmative support for the nonperturbative orbifold equivalence. 
We also comment on subtleties related to domain walls. 

The rest of this paper is structured as follows.
In Section~\ref{sec:SU(n)2_bifundamental}, we introduce the QCD(BF) and review some basics and its conjectured phase diagram.
Before proceeding to the main topic, in Section~\ref{sec:global_inconsistency}, we revisit the 't~Hooft anomalies and global inconsistencies and derive a new constraint on the phase diagram.
In Section~\ref{Sec:semiclassics}, we apply the semiclassical framework to the QCD(BF) via $\mathbb{R}^2 \times T^2$ compactification with 't Hooft flux.
We illustrate how the dilute gas approximation of 2d effective theory gives a vacuum structure, and we obtain the QCD(BF) phase diagram.
The obtained phase diagrams are consistent with the conjectured ones, and concrete examples of phase diagrams at several parameter setups are also shown at the end of this section.
Section~\ref{sec:discussion} includes analytical remarks on domain walls and detailed discussions on the orbifold equivalence.
Section \ref{sec:conclusion} presents a summary of our results, followed by a discussion and an outlook for future research.

\section{Review of \texorpdfstring{$SU(N)\times SU(N)$}{SU(N)xSU(N)} bifundamental QCD}\label{sec:SU(n)2_bifundamental}

In this section, we review the properties of the bifundamental QCD, or QCD(BF).
In Sections \ref{sec:model} and \ref{sec:symmetry}, we introduce the action and symmetries of the $SU(N)_1\times SU(N)_2$ QCD(BF).
In Section~\ref{sec:conjectured_phase_diagram}, we review the conjectured phase diagram based on anomalies and global inconsistencies~\cite{Tanizaki:2017bam} combined with the analysis of various limits~\cite{Karasik:2019bxn}.

\subsection{The model}
\label{sec:model}
QCD(BF) is a $4$-dimensional gauge theory with the gauge group $SU(N)_1\times SU(N)_2$ coupled to the Dirac fermion in the bifundamental representation $(\Box, \overline{\Box})$. 
The classical action is given by
\begin{align}
S=&  \frac{1}{g_1^2} \int  |f_1|^2 +  \frac{1}{g_2^2} \int |f_2|^2 +\int \operatorname{tr} 
 \overline{\Psi}(\slashed{D}+m)\Psi \nonumber\\
&+{\im\theta_1 \over 8\pi^2}\int \operatorname{tr}(f_1\wedge f_1)+{\im\theta_2\over 8\pi^2}\int \operatorname{tr} (f_2\wedge f_2), 
\label{eq:SU(n)_bifundamental_action}    
\end{align}
with the following notations.
$f_i=\diff a_i+\im a_i\wedge a_i$ denotes the gauge field strength of the $SU(N)_i$ gauge field $a_i$ with $i=1,2$, and we introduce the shorthand notation $|f|^2 := \operatorname{tr}(f \wedge \star f)$ for the gauge kinetic terms. The bifundamental fermion $\Psi$ is an $N \times N$ matrix-valued Dirac spinor.
The gauge transformation $(U_1,U_2)\in SU(N)_1\times SU(N)_2$ acts on $\Psi$ and $a_i$ as 
\begin{equation}
    \Psi\mapsto U_1\Psi U_2^{\dagger},\quad 
    a_i\mapsto U_i a_i U_i^{\dagger} - \im\, U_i\diff U_i^{\dagger}\quad (i=1,2),
\end{equation}
and the covariant derivative $\slashed{D}\Psi$ takes the form of
\begin{align}
    \slashed{D}\Psi=\gamma^{\mu}(\partial_{\mu}\Psi+\im\, a_{1\mu}\Psi-\im\, \Psi a_{2\mu}). 
\label{eq:bifund_covariant_derivative}
\end{align}
We can assume $m>0$ without loss of generality because the phase of $m$ is equivalent to the redefinition of $\theta_1+\theta_2$ due to the chiral anomaly.

\subsection{Symmetry, anomaly, and global inconsistency}
\label{sec:symmetry}

The 0-form gauge and global symmetries of this model (at general parameters) are given by
\begin{align}
    G^{[0], \mathrm{gauge+global}} = \frac{SU(N)_{1,\mathrm{gauge}} \times SU(N)_{2,\mathrm{gauge}} \times U(1)_V}{\mathbb{Z}_N \times \mathbb{Z}_N}, 
    \label{eq:total_symmetry}
\end{align}
where $U(1)_V$ is the phase rotation of the fermionic field 
\begin{align}
    \Psi\mapsto \mathrm{e}^{\im\phi}\Psi, \qquad \overline{\Psi}\mapsto \mathrm{e}^{-\im\phi}\overline{\Psi},
\end{align}
Due to the overlaps of $\mathbb{Z}_N \subset SU(N)_{1,\mathrm{gauge}}$, $\mathbb{Z}_N \subset SU(N)_{2,\mathrm{gauge}}$, and $\mathbb{Z}_N \subset U(1)_V$, we need to divide the numerator by $\mathbb{Z}_N \times \mathbb{Z}_N$ to obtain the symmetry group with the faithful representation.
These quotients become important when we discuss 't Hooft anomaly in the next section.

The following symmetries are known to exist at special parameters:
\begin{itemize}
    \item \textbf{Chiral symmetry}
    
    In the massless case ($m = 0$), the classical action~\eqref{eq:SU(n)_bifundamental_action} enjoys the chiral symmetry $U(1)_A$:
\begin{align}
    \Psi\mapsto \mathrm{e}^{\im \gamma_5 \phi}\Psi, \qquad \overline{\Psi}\mapsto \overline{\Psi} \mathrm{e}^{\im \gamma_5 \phi}.
\end{align}
    Due to the Adler-Bell-Jackiw (ABJ) anomaly, this is explicitly broken down to its $\mathbb{Z}_{2N}$ subgroup. The symmetry of the massless QCD(BF) becomes
\begin{align}
    G^{[0], ~\mathrm{gauge+global}}_{\mathrm{massless}} = \frac{SU(N)_{1,\mathrm{gauge}} \times SU(N)_{2,\mathrm{gauge}} \times U(1)_V \times (\mathbb{Z}_{2N})_{\mathrm{chiral}}}{\mathbb{Z}_N \times \mathbb{Z}_N \times \mathbb{Z}_2}. \label{eq:massless_full_symmetry}
\end{align}
Note that $\mathbb{Z}_2$ in the denominator is introduced because $\mathbb{Z}_2$ subgroup is identical in $U(1)_V$ and $(\mathbb{Z}_{2N})_{\mathrm{chiral}}$.
\item \textbf{$\mathrm{CP}$ symmetry}

The $\mathrm{CP}$ (or time-reversal) symmetry exists when $\theta_1 \in \pi \mathbb{Z}$ and $\theta_2 \in \pi \mathbb{Z}$.

\item \textbf{$\mathbb{Z}_2$ exchange symmetry}

If $g_1 = g_2$ (i.e. the dynamical scales are the same) and $\theta_1 = \theta_2$, there is the $\mathbb{Z}_2$ exchange symmetry:
\begin{align}
    a_1 \mapsto -a_2^t,\qquad \Psi\mapsto {\Psi}^t
    \label{eq:exchangesymmetry}
\end{align}
\end{itemize}

In addition to the above 0-form symmetries, this model has the $\mathbb{Z}_N$ 1-form symmetry
\begin{align}
    G^{[1]} = \mathbb{Z}_N^{[1]}
    \label{eq:1formSymmetry}
\end{align}
acting on the Wilson loop of $SU(N)_1$ and $SU(N)_2$ in the common way. 
Combined with Eqs.~\eqref{eq:total_symmetry} and~\eqref{eq:1formSymmetry}, the global symmetry at generic couplings is given by 
\begin{equation}
    G^{\mathrm{global}}=\frac{U(1)_V}{\mathbb{Z}_N}\times \mathbb{Z}_N^{[1]}. 
    \label{eq:globalSymmetry}
\end{equation}
In the absence of the bifundamental fermion (i.e. $m\to \infty$), the model is reduced to two decoupled $SU(N)$ Yang-Mills theories, where the 1-form center symmetry becomes $\mathbb{Z}_N^{[1]} \times \mathbb{Z}_N^{[1]}$.
The bifundamental fermion breaks this symmetry into its diagonal subgroup $\mathbb{Z}_N^{[1]}$ since it is invariant only under the diagonal center, but the potential appearance of the extra $1$-form symmetry as $m\to \infty$ is built in as the quotient structure $U(1)_V/\mathbb{Z}_N$ in~\eqref{eq:globalSymmetry}.

As is now well known, the $4$-dimensional $SU(N)$ Yang-Mills theory has the mixed 't~Hooft anomaly between its $\theta$-periodicity and the $\mathbb{Z}_N^{[1]}$ symmetry~\cite{Gaiotto:2017yup}. 
Similarly, we can find the anomaly and global inconsistencies between $\mathbb{Z}_N^{[1]}$ symmetry and $2 \pi$ shift of $\theta_1$ or $\theta_2$ for QCD(BF)~\cite{Tanizaki:2017bam}. 
Under the presence of the background two-form gauge fields $B$ for $\mathbb{Z}_N^{[1]}$, the $\theta$ periodicity is anomalously extended as 
\begin{align}
    Z_{\theta_1+2\pi,\theta_2} [B] &=  Z_{\theta_1,\theta_2+2\pi} [B] = \exp\left(\frac{\im N}{4 \pi} \int B \wedge B\right) Z_{\theta_1,\theta_2} [B] ,  
    \label{eq:anomaly_previousliterature}
\end{align}
where $Z_{\theta_1,\theta_2}[B]$ is the partition function coupled with $B$. The detailed explanation can be found in Section~\ref{sec:global_inconsistency}. 
The anomalous phase breaks the periodicity either in $\theta_1$ or $\theta_2$, leading to generalized anomalies and/or global inconsistency. 
For example, this rules out the possibility that the vacuum at $(\theta_1,\theta_2) =(0,0)$ and the vacuum at $(\theta_1,\theta_2) =(2\pi,0)$ are the same trivial vacuum.
Thus, one of the following must be true:
the vacuum itself is nontrivial such as deconfined states, or there is a quantum phase transition separating them.
This sort of constraint provides valuable information on the possible phase diagrams.
Note that the global inconsistency arises in the form of ${\im N (\Delta\theta_1 + \Delta\theta_2) \over 8\pi^2}\int (B \wedge B)$, which depends only on the $2\pi$-shifts of $\theta_+ = \theta_1 + \theta_2$.
Within this simple consideration of $\mathbb{Z}_N^{[1]}$ symmetry, no global inconsistency has been found on $\theta_- = \theta_1 - \theta_2$ direction.
For details on possible phase diagrams constrained by this anomaly/global inconsistency, see Ref.~\cite{Tanizaki:2017bam}.

Incidentally, if the fermion is massless, this $B\wedge B$ term shows the mixed 't Hooft anomaly between the $\mathbb{Z}_N^{[1]}$ symmetry and $( \mathbb{Z}_{2N})_{\mathrm{chiral}}$, since the chiral symmetry $(\mathbb{Z}_{2N})_{\mathrm{chiral}}$ is equivalent to a $2\pi$ shift in $\theta_1+\theta_2$ through the ABJ anomaly.

\subsection{Conjectured phase diagram}
\label{sec:conjectured_phase_diagram}

\begin{figure}[t]
\centering
\begin{minipage}{.47\textwidth}
\subfloat[$m \neq 0$]{
\includegraphics[scale=0.5]{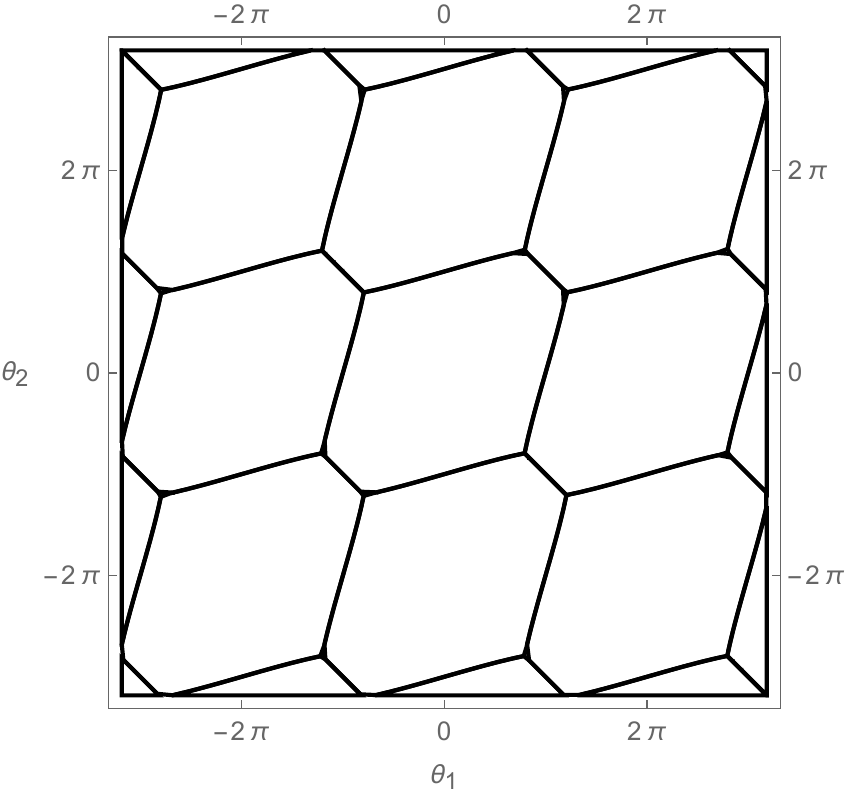}
\label{fig:conjectured_massive}
}\end{minipage}\quad
\begin{minipage}{.47\textwidth}
\subfloat[$m = 0$]{\includegraphics[scale=0.5]{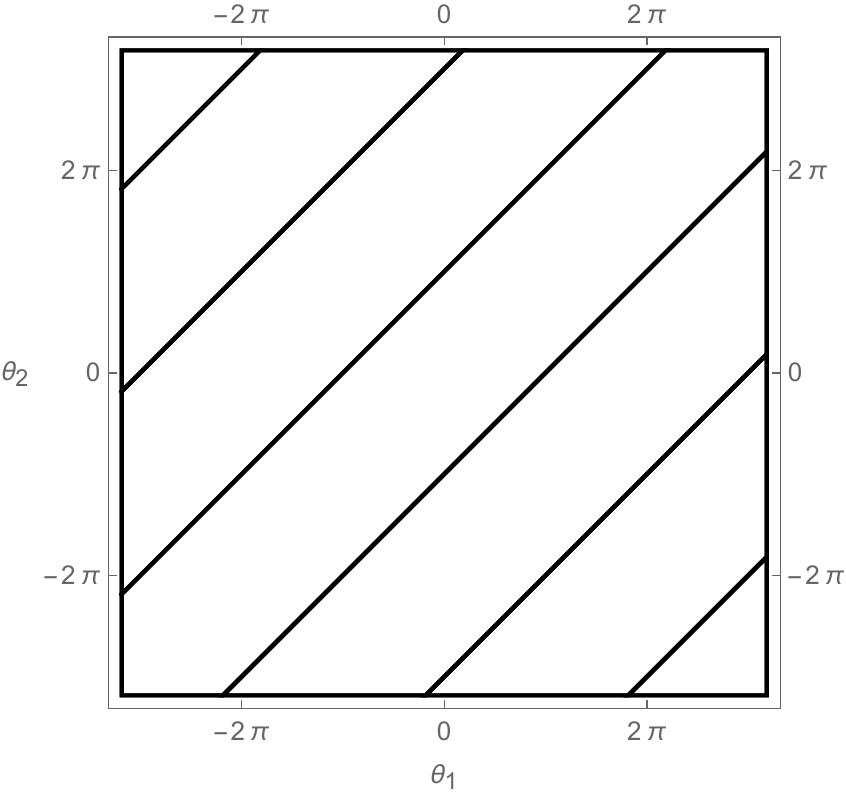}
\label{fig:conjectured_massless}}
\end{minipage}
\caption{Schematic pictures of the conjectured  phase diagrams of QCD(BF) in the $\theta_1$-$\theta_2$ plane. 
(a) Massive case $m \neq 0$: The vacuum is unique and gapped in each phase, and they are separated by the $1$st-order quantum phase transitions. 
(b) Massless case $m = 0$: In each phase, the vacuum is $N$-fold degenerate due to the spontaneous breakdown of the chiral symmetry $(\mathbb{Z}_{2N})_{\mathrm{chiral}} \xrightarrow{\text{SSB}} \mathbb{Z}_2$. 
On the phase boundary, the vacuum is $2N$-fold degenerate.}
\label{fig:conjectured_phase_boundary}
\end{figure}

The constraints by the anomalies and global inconsistencies themselves are not strong enough to determine the phase diagram, and there are several possible scenarios satisfying them. 
Karasik and Komargodski~\cite{Karasik:2019bxn} considered various limits, such as $\Lambda_{2}, m \ll \Lambda_{1}$ and $m \gg  \Lambda_{1,2}$, where $\Lambda_i$ is the dynamical scale of the $SU(N)_i$ gauge group, and proposed a consistent and plausible phase diagram of the QCD(BF).
The conjectured phase diagram of Ref.~\cite{Karasik:2019bxn} is shown in Fig.~\ref{fig:conjectured_phase_boundary}, which is indeed one of the minimal scenarios satisfying the anomaly matching and global inconsistency simultaneously~\cite{Tanizaki:2017bam}:
\begin{itemize}
    \item Massive case $m \neq 0$ (Fig.~\ref{fig:conjectured_massive}).

The vacuum is gapped and unique in each phase, and on the phase boundary, the vacuum is two-fold degenerate. 
This phase diagram is consistent with the global inconsistency above, e.g., $(\theta_1, \theta_2) = (0,0)$ and $(\theta_1, \theta_2) = (2\pi,2\pi)$ are separated by a phase transition line.

    \item Massless case $m=0$ (Fig.~\ref{fig:conjectured_massless}).

In each phase, the vacuum is $N$-fold degenerate, and on the phase boundary, the vacuum is $2N$-fold degenerate.
The anomaly between $\mathbb{Z}_N^{[1]}$ and $( \mathbb{Z}_{2N})_{\mathrm{chiral}}$ is matched by the spontaneously breaking, $( \mathbb{Z}_{2N})_{\mathrm{chiral}}\xrightarrow{\text{SSB}}\mathbb{Z}_2$, leading to $N$-fold degeneracy.
\end{itemize}

The main goal of this paper is to check the validity of these conjectured phase diagrams from totally independent computations. 
Our analysis is based on the recently-proposed semiclassical description for confining vacuum~\cite{Tanizaki:2022ngt, Tanizaki:2022plm}, which will be discussed in Section~\ref{Sec:semiclassics}, and we will obtain the phase diagrams by an explicit computation starting from the QCD(BF) Lagrangian within the given semiclassical setup.

\section{Anomaly and global inconsistency revisited}
\label{sec:global_inconsistency}

In this section, let us reconsider the computations of 't Hooft anomalies and global inconsistencies of QCD(BF).
In addition to the $\mathbb{Z}_N^{[1]}$ background, we introduce the gauge background of $U(1)_V$ with a careful treatment for the quotient (\ref{eq:total_symmetry}).

\subsection{Anomaly and global inconsistency related to the \texorpdfstring{$\theta_{1,2}$}{theta{1,2}} periodicity}

To introduce the gauge background for the 0-form symmetry (\ref{eq:total_symmetry}), we promote the original $SU(N) \times SU(N)$ bundle to the $G^{[0], \mathrm{gauge+global}}$ bundle.
Here, it is important to notice that a $G^{[0], \mathrm{gauge+global}}$ bundle need not to be an $SU(N)_1 \times SU(N)_2 \times U(1)_V$ bundle.
The $\mathbb{Z}_N$ quotients of $G^{[0], \mathrm{gauge+global}}$ relax the cocycle condition of $SU(N) \times SU(N) \times U(1)$ bundle up to $\mathbb{Z}_N \times \mathbb{Z}_N$.
Such deviations from the $SU(N) \times SU(N) \times U(1)$-cocycle condition can be represented as co-dimension-2 defects. 

The convenient way to treat it is to use the background $2$-form gauge fields, and we here follow Refs.~\cite{Kapustin:2014gua}. 
We introduce the $\mathbb{Z}_N$ two-form gauge fields as pairs of $U(1)$ $2$-form and $1$-form gauge fields $(B_i,C_i)$ with the constraints,
\begin{equation}
    NB_i=\diff C_i. 
\end{equation}
We then locally extend the $SU(N)_i$ gauge fields to the $U(N)$ gauge fields, 
\begin{equation}
    \widetilde{a}_i=a_i+\frac{1}{N}C_i \bm{1}_N \quad (\text{locally}),
\end{equation}
which actually means $\tr(\widetilde{a}_i)=C_i$. The corresponding $U(N)$ field strengths are given by $\widetilde{f}_i=\diff \widetilde{a}_i+\im \widetilde{a}_i\wedge \widetilde{a}_i$. 
We also introduce the $U(1)_V$ gauge field $A$, so that the covariant derivative now takes the form, 
\begin{equation}
    \slashed{D} \Psi \Rightarrow \widetilde{\slashed{D}}[A]\Psi=\gamma^{\mu}\left(\partial_\mu \Psi+\im\, \widetilde{a}_1 \Psi-\im\,\Psi \widetilde{a}_2+\im\,A\Psi\right).
    \label{eq:gaugedDirac}
\end{equation}
As a price of the local extension from $SU(N)$ to $U(N)$, we need to postulate the $U(1)$ $1$-form gauge invariance appropriately~\cite{Kapustin:2014gua}. 
The $1$-form gauge transformation is given by 
\begin{equation}
    B_1\mapsto B_1+\diff \Lambda_1,\quad B_2\mapsto B_2+\diff \Lambda_2, 
\end{equation}
where the gauge parameters $\Lambda_{1,2}$ are $U(1)$ $1$-form gauge fields. 
The constraint requires that 
\begin{equation}
    \widetilde{a}_i\mapsto \widetilde{a}_i+\Lambda_i \bm{1}_N,
\end{equation}
and thus $\widetilde{f}_i\mapsto \widetilde{f}_i+\diff \Lambda_i \bm{1}_N$. 
We also require the $1$-form gauge invariance of the covariant derivative, which determines the transformation law of $A$:
\begin{equation}
    A\mapsto A-\Lambda_1+\Lambda_2. 
\end{equation}
These are the basic ingredients for gauging of $G^{\mathrm{global}}$ in \eqref{eq:globalSymmetry}, but it is convenient to rotate the basis of the $\mathbb{Z}_N$ two-form gauge fields. Let us denote 
\begin{equation}
    B_1=B, \quad 
    B_2=B+B_A,
\end{equation}
then the $1$-form gauge-invariant field strengths are given by the following replacements~\cite{Sulejmanpasic:2020zfs},
\begin{align}
    & f_1 \Rightarrow \widetilde{f}_1 - B_1 \bm{1}_N=\widetilde{f}_1-B\bm{1}_N \notag \\
    & f_2 \Rightarrow \widetilde{f}_2 - B_2 \bm{1}_N = \widetilde{f}_2-(B+B_A)\bm{1}_N \notag \\
    & \diff A \Rightarrow  \diff A+B_1-B_2= \diff A - B_A.
    \label{eq:couplingsBackgroundFields}
\end{align}
In this rotated basis, $B$ denotes the background gauge field for the $\mathbb{Z}_N^{[1]}$ symmetry in \eqref{eq:globalSymmetry}. 
To summarize, the gauge backgrounds of the total global symmetries \eqref{eq:globalSymmetry} are represented by 
\begin{itemize}
    \item $B$: $\mathbb{Z}_N$ 2-form gauge field ($\mathbb{Z}_N^{[1]}$ background),
    \item $A$: $U(1)_V$ 1-form gauge field,
    \item $B_A$: $\mathbb{Z}_N$ 2-form gauge field ($\mathbb{Z}_N$ quotient of $U(1)_V/\mathbb{Z}_N$).
\end{itemize}

Now we can readily evaluate the responses under the $2 \pi$ shifts of $(\theta_1, \theta_2)$ in the presence of the gauge backgrounds.
Let $Z_{\theta_1,\theta_2} [A, B_A, B]$ denote the partition function,
\begin{align}
    Z_{\theta_1,\theta_2} [A, B_A, B] := \int \mathcal{D}\widetilde{a}_1 \mathcal{D}\widetilde{a}_2 \mathcal{D} \overline{\Psi} \mathcal{D} \Psi ~\rme^{-S[A, B_A, B]} ,
\end{align}
with the following action
\begin{align}
    S[A, B_A, B] &:=   \frac{1}{g_1^2} \int  |\widetilde{f}_1-B|^2 +  \frac{1}{g_2^2} \int |\widetilde{f}_2-B-B_A|^2 +\int \operatorname{tr} \overline{\Psi}(\widetilde{\slashed{D}}[A]+m)\Psi \nonumber\\
&+{\im\theta_1 \over 8\pi^2}\int \operatorname{tr}\left[(\widetilde{f}_1-B)^2\right]+{\im\theta_2\over 8\pi^2}\int \operatorname{tr} \left[(\widetilde{f}_2-B-B_A)^2\right].
\label{eq:action_with_background}  
\end{align}
In the presence of these backgrounds, the $\theta$-angle periodicity is modified, and this is the origin of global inconsistencies:
\begin{align}
    Z_{\theta_1+2\pi,\theta_2} [A, B_A, B] &= \exp\left(\frac{\im N}{4 \pi} \int B \wedge B\right) Z_{\theta_1,\theta_2} [A, B_A, B] , \notag \\
    Z_{\theta_1,\theta_2+2\pi} [A, B_A, B] &=  \exp\left(\frac{\im N}{4 \pi} \int (B+B_A) \wedge (B+B_A)\right) Z_{\theta_1,\theta_2} [A, B_A, B] .
    \label{eq:anomalyQCDBF}
\end{align}
Let us discuss the consequence of this anomaly/global inconsistency. 
Assume that the system has a unique, gapped, and confining ground state at generic $\theta_1, \theta_2$, then the partition function with the background gauge field is well-described by its topological action:
\begin{equation}
    Z_{\theta_1,\theta_2}[A,B_A,B]\simeq \exp\left(\frac{\im N k}{4\pi}\int B\wedge B+\frac{\im N \ell}{2\pi}\int B\wedge (\diff A-B_A)+\frac{\im N^2 \phi}{8\pi^2}\int (\diff A-B_A)^2\right). 
    \label{eq:SPT}
\end{equation}
Here, $k\sim k+N$ and $\ell\sim \ell+N$ are discrete labels, and $\phi$ is a $2\pi$-periodic continuous parameter. 
The discrete parameters are robust under continuous deformations, and thus they are useful to distinguish vacua as quantum phases of matter. 
Therefore, let us pay attention to the discrete labels $(k,\ell)$. 
The anomaly~\eqref{eq:anomalyQCDBF} shows that these labels should be changed by the shift of the $\theta$ angles:
\begin{align}
    (\theta_1,\theta_2)\mapsto (\theta_1+2\pi n_1,\theta_2+2\pi n_2)\Rightarrow
    (k,\ell)\mapsto (k+n_1+n_2, \ell-n_2). 
\end{align}
As long as $(k,\ell)\not = (k+n_1+n_2,\ell-n_2)$ mod $N$, these two confining vacua should be separated by quantum phase transitions, or the system should be deconfined.

If we turn off $\diff A$ and $B_A$, we can focus on the anomaly that only involves the $\mathbb{Z}_N^{[1]}$ symmetry, and this is done in Ref.~\cite{Tanizaki:2017bam}. 
However, the anomaly related to $\theta_-=\theta_1-\theta_2$ does not appear as we have discussed in Section~\ref{sec:symmetry}. 
In the above discussion, the label $k$ depends only on $n_1+n_2$, and it cannot distinguish different $(n_1,n_2)$ if its sum is the same. 
Turning on the $U(1)_V/\mathbb{Z}_N$ gauge field, $N(\diff A-B_A)$, gives a new constraint on the $\theta_-$ dependence:
\begin{align}
    Z_{\theta_1+2\pi,\theta_2-2\pi} [A, B_A, B] &= \exp\left(-\frac{\im N}{2 \pi} \int B \wedge B_A - \frac{\im N}{4 \pi} \int B_A \wedge B_A\right) Z_{\theta_1,\theta_2} [A, B_A, B].
\end{align}
The second term can be written as $\frac{\im N}{4\pi} \int (\diff A - B_A)^2$ (mod $2\pi$) and thus this anomalous phase can be eliminated by a shift of the continuous $\theta$ angle, which does not provide a useful constraint on the phase diagram. 
However, the cross term $-\frac{\im N}{2\pi}\int B \wedge B_A$, which can be written as $\frac{\im N}{2\pi}\int B\wedge (\diff A-B_A)$, gives a shift of the discrete theta term, 
and thus we can regard it as the mixed anomaly between $U(1)_V/\mathbb{Z}_N$, $\mathbb{Z}_N^{[1]}$, and the $\theta_-$ periodicity (see Refs.~\cite{Shimizu:2017asf, Gaiotto:2017tne, Tanizaki:2017mtm, Tanizaki:2018wtg, Yonekura:2019vyz, Anber:2019nze, Morikawa:2022liz} for related anomalies).
This anomaly matching/global inconsistency requires that phases at $\theta_- = 0$ and $\theta_-= 2\pi$ are separated by at least one phase transition (or these phases are nontrivial). 
This rules out a possibility that $(\theta_1,\theta_2) = (0,0)$ and $(\theta_1,\theta_2) = (2\pi,-2\pi)$ are connected by a gapped single vacuum, which was suggested as one of the possible scenarios in Ref.~\cite{Tanizaki:2017bam}. 
We note that the conjectured phase diagram in Fig.~\ref{fig:conjectured_phase_boundary} is still consistent with this new constraint.

\subsection{Anomaly related to the chiral symmetry \texorpdfstring{$(\mathbb{Z}_{2N})_{\mathrm{chiral}}$}{(Z{2N})chiral}}

We now consider the massless case, $m=0$, and let us then calculate the anomaly between $\frac{U(1)_V}{\mathbb{Z}_N}\times \mathbb{Z}_N^{[1]}$ and the chiral symmetry, $(\mathbb{Z}_{2N})_{\mathrm{chiral}}$ (see also Section~4.2 of Ref.~\cite{Sulejmanpasic:2020zfs}).
We again introduce the background gauge field for the vector-like symmetry, $\frac{U(1)_V}{\mathbb{Z}_N}\times \mathbb{Z}_N^{[1]}$, and then the index of the Dirac operator~\eqref{eq:gaugedDirac} is given by 
\begin{align}
    \mathrm{ind}(\widetilde{\slashed{D}}[A])
    &=\int\frac{1}{8\pi^2}\tr\left[(\widetilde{f}_1\otimes \bm{1}_N-\bm{1}_N\otimes\widetilde{f}_2+\diff A \bm{1}_N\otimes \bm{1}_N)^2\right] \notag\\
    &=\int \frac{1}{8\pi^2}\tr\left[\left((\widetilde{f}_1-B)\otimes \bm{1}_N-\bm{1}_N\otimes(\widetilde{f}_2-(B+B_A))+(\diff A-B_A)\bm{1}_N\otimes \bm{1}_N \right)^2\right]\notag\\
    &=\int \frac{1}{8\pi^2}\left(N \tr(\widetilde{f}_1-B)^2 + N \tr(\widetilde{f}_2-B-B_A)^2+N^2(\diff A-B_A)^2\right). 
\end{align}
Under the discrete chiral transformation, the fermion integration measure changes as 
\begin{align}
    \Diff \overline{\Psi} \Diff \Psi & \mapsto \exp\left(\frac{2\pi \im}{N}\mathrm{ind}(\widetilde{\slashed{D}}[A])\right) \Diff \overline{\Psi}\Diff \Psi \notag\\
    &= \exp\left(\im\int\left\{-\frac{2N}{4\pi}B^2+\frac{N}{2\pi}B\wedge (\diff A-B_A)\right\}\right) \Diff \overline{\Psi}\Diff \Psi. 
    \label{eq:chiralanomaly}
\end{align}
Let us again assume the confining vacua, then this anomaly requires some chiral symmetry breaking. If the vacuum were unique and gapped, the partition function would have to be described by \eqref{eq:SPT}, but the chiral transformation acts on its label as 
\begin{equation}
    (k,\ell)\mapsto (k-2, \ell+1). 
\end{equation}
As we have to repeat this operation $N$ times to go back to the original state, the anomaly matching always requires the complete spontaneous breaking of the chiral symmetry, 
\begin{equation}
    (\mathbb{Z}_{2N})_{\mathrm{chiral}}\xrightarrow{\text{SSB}} \mathbb{Z}_2, 
\end{equation}
if the system is gapped and confining~\cite{Sulejmanpasic:2020zfs}. 
Again, even if we only consider the anomaly involving $B$, there is a mixed chiral anomaly and we can conclude some chiral symmetry breaking. 
However, the anomaly involving $B$ is not enough to conclude the complete breakdown of chiral symmetry when $N$ is even, and the mixed anomaly between $\mathbb{Z}_N^{[1]}$, $U(1)_V/\mathbb{Z}_N$, and chiral symmetry plays an important role.

\subsection{Comments on the \texorpdfstring{$\mathbb{Z}_2$}{Z2} exchange symmetry}

As we discussed in Section~\ref{sec:symmetry}, there is the $\mathbb{Z}_2$ exchange symmetry if $g_1=g_2$ and $\theta_1=\theta_2$. 
In Ref.~\cite{Tanizaki:2017bam}, it is argued that the anomaly related to this $\mathbb{Z}_2$ symmetry does not exist, and this symmetry is assumed to be unbroken in the conjectured phase diagram, Fig.~\ref{fig:conjectured_phase_boundary}. 
Let us revisit this point from the viewpoint of low-energy effective action. 

For this purpose, we need to identify how the $\mathbb{Z}_2$ exchange symmetry acts on the background gauge fields. 
The transformation rule~\eqref{eq:exchangesymmetry} is uniquely extended to the background gauge fields, 
\begin{equation}
    (\mathbb{Z}_2)_{\mathrm{exchange}}: B\mapsto -B-B_A,\quad  (A,B_A)\mapsto (A,B_A),
\end{equation}
for invariance of the action. Substituting this transformation property into the effective action~\eqref{eq:SPT}, we obtain that the $\mathbb{Z}_2$ exchange symmetry acts on its label $(k,\ell)$ as 
\begin{equation}
    (k,\ell)\mapsto (k,-k-\ell). 
\end{equation}
If we change the basis of $\mathbb{Z}_N$ labels as 
\begin{equation}
    k=k_1+k_2,\quad \ell=-k_2, 
\end{equation}
then the action of the $\mathbb{Z}_2$ exchange symmetry is $k_1\leftrightarrow k_2$. 
We can find the $\mathbb{Z}_2$ invariant states, such as $(k,\ell)=(0,0), (2,-1), (4,-2),$ etc. ($\Leftrightarrow (k_1,k_2)=(0,0), (1,1), (2,2)$ etc.), and these are confining $\mathbb{Z}_2$-invariant states. 
Let us note, however, that there are many $\mathbb{Z}_2$-paired states, such as $(0,1)\leftrightarrow (0,-1)$, $(1,0)\leftrightarrow (1,-1)$, etc. ($\Leftrightarrow k_1\not=k_2$).
It is a dynamical question if the actual vacuum of QCD(BF) prefers the unbroken or broken $\mathbb{Z}_2$ symmetry, and this is one of the main questions we are going to address in the following.

\section{Semiclassics by anomaly-preserving \texorpdfstring{$T^2$}{T2} compactification}
\label{Sec:semiclassics}

As we have seen in the previous section, the anomaly and global inconsistency give the severe constraints on the possible vacuum structures. 
However, there still exist multiple scenarios for the possible phase diagrams, and we need to have some knowledge on the dynamics to pin down which scenario is more plausible. 

In this section, we study the dynamics of QCD(BF) on the novel semiclassical computations on $\mathbb{R}^2\times T^2$ with the 't~Hooft flux~\cite{Tanizaki:2022ngt, Tanizaki:2022plm}. 
In Section~\ref{sec:methodology}, we introduce this calculable framework by anomaly-preserving $T^2$ compactification. 
In Section~\ref{sec:semiclassics_massless}, 
we compute the phase diagram for the massless case and show that the vacuum just breaks the chiral symmetry as indicated in the conjectured phase diagram, Fig.~\ref{fig:conjectured_massless}.
The calculation in the presence of mass perturbation is presented in Section~\ref{sec:semiclassics_massive}, and this again supports the minimal scenario of the phase diagram, Fig.~\ref{fig:conjectured_massive}.
In Section~\ref{sec:intermediate_m}, we show phase diagrams computed numerically at several concrete parameters and mention phase diagrams with dynamical scale hierarchy.

\subsection{Methodology}
\label{sec:methodology}

Recently, a new calculable framework has been proposed by anomaly-preserving $T^2$ compactification~\cite{Tanizaki:2022ngt, Tanizaki:2022plm} (see also Ref.~\cite{Yamazaki:2017ulc}). 
The biggest advantage is that the theory becomes weakly coupled when the compactification size $L$ is small, suggesting the validity of semiclassical calculations.
Indeed, it has been shown in Ref.~\cite{Tanizaki:2022ngt} that the dilute gas approximation of fractional instantons, or center vortices, successfully describes the confining vacua as long as we insert the suitable 't~Hooft flux.

Of course, this weak-coupling description is applicable only for sufficiently small $L$, and it is still difficult to study the strongly-coupled dynamics in the decompactified limit. 
We here assume the adiabatic continuity, i.e., the absence of phase transition, between the small-$L$ theory and the large-$L$ theory (toward decompactifying limit $L \rightarrow \infty$).
Assuming the adiabatic continuity conjecture, we can propose the qualitative features of the theory on $\mathbb{R}^4$ through calculations of the theory on $\mathbb{R}^2 \times T^2$. Schematically,
\begin{align}
    &\mathcal{T}_{\mathbb{R}^4} ~~~~~~~~~ \simeq ~~~~~~~~~~ \mathcal{T}_{\mathbb{R}^2 \times T^2}   \\
    \mathrm{~(strongly}&\mathrm{~coupled)~} ~~~~~ \mathrm{~(weakly}\mathrm{~coupled)~}, \notag
\end{align}
where $\mathcal{T}_{\mathbb{R}^4}$ and $\mathcal{T}_{\mathbb{R}^2 \times T^2}$ are the theories on $\mathbb{R}^4$ and on $\mathbb{R}^2 \times T^2$, respectively.
Indeed, it has been established that this semiclassical calculation on $\mathbb{R}^2\times T^2$ provides qualitatively consistent results for $SU(N)$ Yang-Mills theory, $\mathcal{N} = 1$ supersymmetric Yang-Mills theory, QCD in the chiral limit \cite{Tanizaki:2022ngt}, and all these results support the adiabatic continuity conjecture for QCD-like theories. 
This method is also applied to study the structures of QCD with two-index quarks~\cite{Tanizaki:2022plm}.
In the following, we take this approach to understand the phase diagram of QCD(BF).

\subsubsection{Anomaly-preserving compactification}
``Anomaly-preserving compactification'' (from $M_4$ to $M_2 \times T^2$) is implemented as follows~\cite{Tanizaki:2022ngt} (see also~\cite{Tanizaki:2017qhf, Yamazaki:2017dra, Dunne:2018hog}).
The massless QCD(BF) has the mixed 't Hooft anomaly~\eqref{eq:chiralanomaly} between $U(1)_V/\mathbb{Z}_N$, $\mathbb{Z}_N^{[1]}$, and $(\mathbb{Z}_{2N})_{\mathrm{chiral}}$, 
and this anomaly is controlled by the following 5d topological action:
\begin{align}
    I_{5d} = \frac{2\pi\cdot 2}{N} \int_{M_5} \frac{2N}{2\pi}A_{\mathrm{chiral}} \wedge \frac{N^2}{8\pi^2}B^2-\frac{2\pi}{N}\int_{M_5} \frac{2N}{2\pi}A_{\mathrm{chiral}}\wedge \frac{N}{2\pi}B\wedge \frac{N}{2\pi}(\diff A-B_A),
\end{align}
where $A_{\mathrm{chiral}}$ is the $(\mathbb{Z}_{2N}^{[0]})_{\mathrm{chiral}}$ background field and $M_5$ is a 5-dimensional manifold with $\partial M_5 = M_4$.

Let us compactify $M_4$ into $M_2 \times T^2$ with the size of torus $L$.
We use coordinates $\Vec{x} = (x_1,x_2)$ for $M_2$ and $(x_3,x_4)$ for $T^2$ with identifying $x_3 \sim x_3 + L$ and $x_4 \sim x_4 + L$.
At this moment, we can introduce the 't Hooft twist on $T^2$ by setting an appropriate boundary condition, labeled by $n_{34} \in \mathbb{Z}_N$, which corresponds to a $\mathbb{Z}_N^{[1]}$-background on $T^2$: $B = \frac{2 \pi n_{34}}{N} \frac{\diff x^3 \wedge \diff x^4}{L^2}$.
This compactification effectively reduces the 4d theory to a 2d theory, and accordingly the 1-form symmetry $\mathbb{Z}_N^{[1]}$ is decomposed into:
\begin{align}
    (\mathbb{Z}_N^{[1]})_{4d} \rightarrow (\mathbb{Z}_N^{[1]})_{2d} \times \mathbb{Z}_N^{[0]} \times \mathbb{Z}_N^{[0]}
\end{align}
In terms of the background gauge field, this decomposition can be expressed as
\begin{align}
    B_{4d} = B_{2d} + A_3 \wedge \frac{\diff x_3}{L} + A_4 \wedge \frac{\diff x_4}{L} + \frac{2 \pi n_{34}}{N} \frac{\diff x^3 \wedge \diff x^4}{L^2}
\end{align}
Here, $B_{2d}$ and $A_3, A_4$ are background gauge fields for $(\mathbb{Z}_{N}^{[1]})_{2d}$ and $\mathbb{Z}_N^{[0]}\times \mathbb{Z}_{N}^{[0]}$, respectively, and they are taken to be independent of $x_3,x_4$. 

Then, the anomaly of the 2d effective theory is described by the 3d action with 2-form $(\mathbb{Z}_N^{[1]})_{2d}$ background $B_{2d}$, 1-form $\mathbb{Z}_N^{[0]} \times \mathbb{Z}_N^{[0]}$ backgrounds $A_3,~A_4$, and the $U(1)_V/\mathbb{Z}_N$ gauge field $(A,B_A)$:\footnote{The anomaly polynomial also contains the terms that involve the $(-1)$-form symmetry coming out of $U(1)_V/\mathbb{Z}_N$, but we here neglect them. }
\begin{align}
    I_{3d} =& \frac{2\pi \cdot 2n_{34}}{N} \int_{M_3} \frac{2N}{2\pi}A_{\mathrm{chiral}} \wedge \frac{N}{2\pi} B_{2d} 
    -\frac{2\pi\cdot 2}{N} \int_{M_3} \frac{2N}{2\pi}A_{\mathrm{chiral}} \wedge \frac{N^2}{(2\pi)^2}(A_3 \wedge A_4)\notag\\
    &-\frac{2\pi n_{34}}{N}\int_{M_3} \frac{2N}{2\pi}A_{\mathrm{chiral}}\wedge \frac{N}{2\pi}(\diff A-B_A). 
\end{align}
With the usual periodic boundary condition ($n_{34} = 0$), the anomaly exists only for $(\mathbb{Z}_{2N}^{[0]})_{\mathrm{chiral}}$ and $\mathbb{Z}_N^{[0]} \times \mathbb{Z}_N^{[0]}$, and the 2d 1-form symmetry $(\mathbb{Z}_N^{[1]})_{2d}$ does not have a mixed anomaly with the chiral symmetry in this case.
Then, the anomaly can be naturally matched by the deconfined phase with the spontaneous breakdown of $\mathbb{Z}_N^{[0]} \times \mathbb{Z}_N^{[0]}$.
On the other hand, in the presence of the 't Hooft flux ($n_{34} \neq 0$), this compactification retains the mixed anomaly between $(\mathbb{Z}_{2N}^{[0]})_{\mathrm{chiral}}$ and $(\mathbb{Z}_N^{[1]})_{2d}$ and also the anomaly between $(\mathbb{Z}_{2N})_{\mathrm{chiral}}$ and $U(1)_V/\mathbb{Z}_N$.
All of these three anomalies can be simultaneously matched by the spontaneous breakdown of chiral symmetry.
This observation motivates us to employ compactification with nontrivial 't Hooft flux (in particular, $n_{34}=1$) to study the confining vacua.\footnote{As in calculations for the fundamental QCD~\cite{Tanizaki:2022ngt}, we could also employ anomaly-preserving compactification with the $U(1)_V/\mathbb{Z}_N$ magnetic flux.
This would be necessary for bifundamental gauge theories with different ranks $SU(N_1)_1\times SU(N_2)_2$, $(N_1 \neq N_2)$. 
Let us leave analyses on different-rank bifundamental gauge theories as a future task.
}

\subsubsection{Semiclassics by center vortices}

In what follows, we use the anomaly-preserving compactification to $\mathbb{R}^2 \times T^2$ with $n_{34} = 1$.
More explicitly, we impose the 't~Hooft twisted boundary conditions on $T^2$~\cite{tHooft:1979rtg}: for the bifundamental fermion $\Psi(x_1,x_2,x_3,x_4)$, 
\begin{align}
    \Psi(x_1,x_2,x_3+L,x_4) &= g_3^{(1)}(x_4) \Psi(x_1,x_2,x_3,x_4) g_3^{(2) \dagger}(x_4) \notag \\ 
    \Psi(x_1,x_2,x_3,x_4+L) &= g_4^{(1)}(x_3) \Psi(x_1,x_2,x_3,x_4) g_4^{(2) \dagger}(x_3) \label{eq:bdy_cond_fermion}
\end{align}
where $(g_3^{(1)}(x_4), g_3^{(2)}(x_4)) \in SU(N)_1 \times SU(N)_2$ and $(g_4^{(1)}(x_3), g_4^{(2)}(x_3)) \in SU(N)_1 \times SU(N)_2$ denote the transition functions of the $SU(N)_1 \times SU(N)_2$ bundle.
In terms of these transition functions of the boundary condition, the minimal $\mathbb{Z}_N^{[1]}$-background on $T^2$, or 't Hooft flux $n_{34} = 1$, means
\begin{align}
    g_3^{(1)\dagger}(L) g_4^{(1)\dagger}(0) = g_4^{(1)\dagger}(L) g_3^{(1)\dagger}(0) \,\rme^{\frac{2 \pi \im}{N}}, \notag \\
    g_3^{(2)\dagger}(L) g_4^{(2)\dagger}(0) = g_4^{(2)\dagger}(L) g_3^{(2)\dagger}(0) \,\rme^{\frac{2 \pi \im}{N}}.
\end{align}
Here we can fix the gauge so that these transition functions are given by shift and clock matrices,
\begin{align}
    g_3^{(1)}(x_4) &= S,~~~ g_4^{(1)}(x_3) = C, \notag \\ 
    g_3^{(2)}(x_4) &= S,~~~ g_4^{(2)}(x_3) = C, \label{eq:transition_fct}
\end{align}
where $C \propto \operatorname{diag}(1,\rme^{\frac{2 \pi \im}{N}}, \cdots,\rme^{\frac{2 \pi \im (N-1)}{N}})$ and $(S)_{ij} \propto \delta_{i+1,j}$.

If we virtually compactify $\mathbb{R}^2$ with 't Hooft flux $n_{12}^{(1)} = 1$ only for $SU(N)_1$, there exists a classical fractional instanton configuration with
\begin{align}
    (Q_{\mathrm{top}}^{(1)}, Q_{\mathrm{top}}^{(2)}) = (1/N,0),~~ S = S_\mathrm{I}^{(1)}/N,
\end{align}
where $Q_{\mathrm{top}}^{(i)}$ denotes the topological charge for the $SU(N)_i$ gauge field and $S_\mathrm{I}$ denotes the instanton action $S_\mathrm{I}^{(1)} = \frac{8 \pi^2}{g^2_1}$.
Similarly, we can get the $(Q_{\mathrm{top}}^{(1)}, Q_{\mathrm{top}}^{(2)}) = (0,1/N)$ fractional instanton with the action $S = S_\mathrm{I}^{(2)}/N = \frac{8 \pi^2}{N g^2_2}$ by the compatification with the flux $n_{12}^{(2)} = 1$.
Here, we assume such a solution exists and survives for the decompactifying limit $T^2 \times T^2 \rightarrow \mathbb{R}^2 \times T^2$, as suggested by numerical works~\cite{Gonzalez-Arroyo:1998hjb, Montero:1999by, Montero:2000pb} (For recent analytical studies, see Refs.~\cite{Anber:2022qsz, Anber:2023sjn}).
In the 2d effective theory on $\mathbb{R}^2$, these fractional instantons can be identified as center vortices.
We then obtain the dilute gas picture generated by $(Q_{\mathrm{top}}^{(1)}, Q_{\mathrm{top}}^{(2)}) = (1/N,0)$ and $(Q_{\mathrm{top}}^{(1)}, Q_{\mathrm{top}}^{(2)}) = (0,1/N)$ fractional instantons and anti-instantons~\cite{Tanizaki:2022ngt}.

Let us first consider the pure $SU(N)\times SU(N)$ Yang-Mills theory without a bifundamental fermion.
The boundary conditions for the gauge fields with the choice (\ref{eq:transition_fct}),
\begin{align}
    a_{1\mu} (x_1,x_2,x_3+L,x_4) &= S a_{1\mu}(x_1,x_2,x_3,x_4) S^\dagger, \notag \\
    a_{1\mu}(x_1,x_2,x_3,x_4+L) &= C a_{1\mu}(x_1,x_2,x_3,x_4) C^\dagger, \notag \\
    a_{2\mu} (x_1,x_2,x_3+L,x_4) &= S a_{2\mu}(x_1,x_2,x_3,x_4) S^\dagger, \notag \\ a_{2\mu}(x_1,x_2,x_3,x_4+L) &= C a_{2\mu}(x_1,x_2,x_3,x_4) C^\dagger, 
    \label{eq:bdy_cond_gauge}
\end{align}
admit no constant mode in $su(N)$, and a perturbative excitation has a gap of order  $O(1/NL)$.
By taking a sufficient small torus $T^2$,
we can neglect these massive Kaluza-Klein modes. 
The Boltzmann weights of center vortices and anti-vortices for $(Q_{\mathrm{top}}^{(1)}, Q_{\mathrm{top}}^{(2)}) = (1/N,0)$ and $(Q_{\mathrm{top}}^{(1)}, Q_{\mathrm{top}}^{(2)}) = (0,1/N)$ are,
\begin{align}
(1/N,0)~\mathrm{vortex}~~
\begin{cases}
    \mathcal{V}_1(x_1,x_2) = K_*^{(1)} \rme^{-S_\mathrm{I}^{(1)}/N + \im \theta_1 /N} \\
    \mathcal{V}_1^* (x_1,x_2) = K_*^{(1)} \rme^{-S_\mathrm{I}^{(1)}/N - \im \theta_1 /N},    
\end{cases} \notag \\
(0,1/N)~\mathrm{vortex}~~
\begin{cases}
    \mathcal{V}_2(x_1,x_2) = K_*^{(2)} \rme^{-S_\mathrm{I}^{(2)}/N + \im \theta_2 /N} \\
    \mathcal{V}_2^* (x_1,x_2) = K_*^{(2)} \rme^{-S_\mathrm{I}^{(2)}/N - \im \theta_2 /N},    
\end{cases}
\end{align}
with some dimensionful weight $K_*^{(1)},~K_*^{(2)} \sim 1/L^2$.
The dilute gas approximation would yield
\begin{align}
    Z_{SU(N)\times SU(N) \mathrm{YM}} &= \left( \sum_{n_1, \bar{n}_1 \geq 0} \frac{(V K_*^{(1)})^{n_1 + \bar{n}_1}}{n_1! \bar{n}_1 !} \rme^{- \frac{(n_1 + \bar{n}_1) S_\mathrm{I}}{N} + \im \theta_1 \frac{n_1 - \bar{n}_1}{N}} \delta_{n_1-\bar{n}_1 \in N \mathbb{Z}}\right) \notag \\
    &~~~~\times \left( \sum_{n_2, \bar{n}_2 \geq 0} \frac{(V K_*^{(2)})^{n_2 + \bar{n}_2}}{n_2! \bar{n}_2 !} \rme^{- \frac{(n_2 + \bar{n}_2) S_\mathrm{I}}{N} + \im \theta_2 \frac{n_2 - \bar{n}_2}{N}} \delta_{n_2-\bar{n}_2 \in N \mathbb{Z}}  \right),
\end{align}
where the Kronecker delta $\delta_{n_i-\bar{n}_i \in N \mathbb{Z}}$ is introduced because the theory on $M_2 \times T^2$ admits only configurations with integer total topological charges $(Q_{\mathrm{top}}^{(1)}, Q_{\mathrm{top}}^{(2)})$, and $V$ is the volume of $M_2$.

Next, let us include the bifundamental fermion $\Psi$.
The boundary condition (\ref{eq:bdy_cond_fermion}) for the fermion $\Psi$ with the choice (\ref{eq:transition_fct}):
\begin{align}
    \Psi(x_1,x_2,x_3+L,x_4) &= S \Psi(x_1,x_2,x_3,x_4) S^\dagger \notag \\ 
    \Psi(x_1,x_2,x_3,x_4+L) &= C \Psi(x_1,x_2,x_3,x_4) C^\dagger \label{eq:fermion_zeromode}
\end{align}
admits one (4d spinor) constant mode on $T^2$: $\Psi(x) = \frac{1}{\sqrt{N} L}1_{N \times N} \psi(x_1,x_2)$.
For the sufficiently small torus $T^2$, the bifundamental fermion is described by the above zero mode in the 2d effective theory:
\begin{align}
    S_{2d}^{\mathrm{fermion}}[\psi, \bar{\psi}] = \int \bar{\psi} (\slashed{\partial}_{2d}+m) \psi  
\end{align}
where $\slashed{\partial}_{2d} = \gamma^1 \partial_1 + \gamma^2 \partial_2$ is the $2$d Dirac operator with the $4$d gamma matrices.
The $2$d fermion picture will be valid for $|m| \ll 1/(NL)$. For $|m| \gtrsim 1/(NL)$, the 2d fermion would be as heavy as perturbative gluons or other Kaluza-Klein modes, which should be integrated out for sufficiently small $L$.

To incorporate the fermion into the dilute gas approximation, we need to evaluate the effect of the center vortex on the fermion.
Minimally, such effects are qualitatively estimated by counting fermionic zeromodes around the center vortex.
We have $(Q_{\mathrm{top}}^{(1)}, Q_{\mathrm{top}}^{(2)}) = (\pm 1/N,0)$ vortex and $(Q_{\mathrm{top}}^{(1)}, Q_{\mathrm{top}}^{(2)}) = (0,\pm 1/N)$ vortex with the same index
\begin{align}
     2 \operatorname{ind} (\slashed{D}) = 2N(|Q_{\mathrm{top}}^{(1)}| + |Q_{\mathrm{top}}^{(2)}|) = 2.
\end{align}
We can then determine the vertex operator of these center vortices qualitatively:
\begin{align}
(\pm 1/N,0)~\mathrm{vortex}~~
\begin{cases}
    \mathcal{V}_1(\Vec{x}) = -K^{(1)} \rme^{-S_\mathrm{I}^{(1)}/N + \im \theta_1 /N} (\bar{\psi}_L \psi_R)(\Vec{x}) \\
    \mathcal{V}_1^* (\Vec{x}) = -K^{(1)} \rme^{-S_\mathrm{I}^{(1)}/N - \im \theta_1 /N} (\bar{\psi}_R \psi_L)(\Vec{x}), 
\end{cases} \notag \\
(0,\pm 1/N)~\mathrm{vortex}~~
\begin{cases}
    \mathcal{V}_2(\Vec{x}) = -K^{(2)} \rme^{-S_\mathrm{I}^{(2)}/N + \im \theta_2 /N} (\bar{\psi}_L \psi_R)(\Vec{x})\\
    \mathcal{V}_2^* (\Vec{x}) = -K^{(2)} \rme^{-S_\mathrm{I}^{(2)}/N - \im \theta_2 /N} (\bar{\psi}_R \psi_L)(\Vec{x}),  \label{eq:vertex_oper}
\end{cases}
\end{align}
where $K^{(1)},~K^{(2)} \sim 1/L$ is a dimensionful positive constant and $\Vec{x} = (x_1,x_2)$. Here, the sign of $K^{(1)}$ and $K^{(2)}$ is rather a convention of theta angles or phases of fermion fields, and we take $K^{(1)},~K^{(2)} >0$ so that the fermion has a positive mass at $(\theta_1,\theta_2) = (0,0)$ in the trivial sector $(k_1,k_2) = (0,0)$.

We can thus obtain the partition function by the dilute gas approximation:
\begin{align}
    Z_{\theta_1,\theta_2} &= \int \mathcal{D}\psi \mathcal{D}\Bar{\psi}~\rme^{-S_{2d}^{\mathrm{fermion}} [\psi, \bar{\psi}]} \notag \\
    &~~~~\times \left[ \sum_{n_1, \bar{n}_1 \geq 0} \frac{1}{n_1! \bar{n}_1 !} \left( \int \rmd^2\Vec{x} ~\mathcal{V}_1(\Vec{x}) \right)^{n_1} \left( \int \rmd^2\Vec{x} ~\mathcal{V}_1^*(\Vec{x}) \right)^{\bar{n}_1} \delta_{n_1-\bar{n}_1 \in N \mathbb{Z}}\right] \notag \\
    &~~~~\times \left[ \sum_{n_2, \bar{n}_2 \geq 0} \frac{1}{n_2! \bar{n}_2 !} \left( \int \rmd^2\Vec{x} ~\mathcal{V}_2(\Vec{x}) \right)^{n_2} \left( \int \rmd^2\Vec{x} ~\mathcal{V}_2^*(\Vec{x}) \right)^{\bar{n}_2} \delta_{n_2-\bar{n}_2 \in N \mathbb{Z}}\right]
    \notag\\
    &= \frac{1}{N^2} \sum_{k_1, k_2 \in \mathbb{Z}_N} \int \mathcal{D}\psi \mathcal{D}\Bar{\psi} \rme^{-S_{2d}^{(k_1,k_2)}[\psi, \bar{\psi}]}, 
    \label{eq:dilute_gas_paritition_fct}
\end{align}
where 
\begin{align}
    S_{2d}^{(k_1,k_2)}[\psi, \bar{\psi}] &:= S_{2d}^{\mathrm{fermion}}[\psi, \bar{\psi}] \notag\\
    &\qquad - \int \rmd^2\Vec{x} ~\left[ \rme^{- \frac{2\pi \im k_1}{N} }\mathcal{V}_1(\Vec{x}) + \rme^{ \frac{2\pi \im k_1}{N}}  \mathcal{V}_1^*(\Vec{x}) +  \rme^{- \frac{2\pi \im k_2}{N} }\mathcal{V}_2(\Vec{x}) + \rme^{\frac{2\pi \im k_2}{N}}  \mathcal{V}_2^*(\Vec{x}) 
    \right] \notag \\
    &= S_{2d}^{\mathrm{fermion}}[\psi, \bar{\psi}] + \int \rmd^2\Vec{x} ~\Bigl[  M_{k_1,k_2} (\bar{\psi}_L \psi_R)(\Vec{x}) + M_{k_1,k_2}^* (\bar{\psi}_R \psi_L)(\Vec{x})  \Bigr], \label{eq:sector_action}  \\
    M_{k_1,k_2} &:=  K^{(1)} \rme^{-S_\mathrm{I}^{(1)}/N + \im (\theta_1-2\pi k_1) /N} + K^{(2)} \rme^{-S_\mathrm{I}^{(2)}/N + \im (\theta_2-2\pi k_2) /N}. \label{eq:effective_mass_massless}
\end{align}
This expression indicates that the 2d effective theory has $N^2$ semiclassical vacua $\{ \ket{k_1,k_2}\}$ (as in $SU(N) \times SU(N)$ pure Yang-Mills theory) and, in each semiclassical vacuum, the fermion with mass $M_{k_1,k_2}+m$ fluctuates.
In the following sections, we will see how the fermion chooses ground states and draw the phase diagrams.

Beforehand, let us mention an interpretation of the $N^2$ semiclassical vacua $\ket{k_1,k_2}$ with $k_1,k_2 \in \mathbb{Z}_N$ by a further compactification (For details in $SU(N)$ pure Yang-Mills theory, see Section 2.5 of \cite{Tanizaki:2022ngt}).
One can further compactify $\mathbb{R}^2$ into $\mathbb{R}\times S^1_A$ with the periodic boundary condition with radius $L_A$ ($\Lambda^{-1} \gg L_A \gg L$), namely,
\begin{align}
    \mathbb{R}^4 \rightarrow \mathbb{R} \times S^1_A \times (S^1_B \times S^1_C) \label{eq:R_S1_3_setup}
\end{align}
 with the 't Hooft flux for $(S^1_B \times S^1_C)$.
 In this setup, we have $N^2$ classical vacua $\ket{m_1,m_2}$ with $m_1,m_2 \in \mathbb{Z}_N$, labeled by Polyakov loops:
 \begin{align}
     P_{a_1}(S^1_A) =  \rme^{\frac{2 \pi\im m_1}{N}} \textbf{1}_{N \times N},~~ P_{a_2}(S^1_A) =  \rme^{\frac{2 \pi\im m_2}{N}} \textbf{1}_{N \times N}
 \end{align}
Here, a fractional instanton describes the tunneling processes $\ket{m_1,m_2} \rightarrow \ket{m_1 \pm 1,m_2}$ or $\ket{m_1,m_2} \rightarrow \ket{m_1 ,m_2\pm 1}$ accompanying fermion zeromodes.
Schematically, one fractional instanton yields
\begin{align}
    &\braket{m_1',m_2'|e^{-T \hat{H}}|m_1,m_2} \sim \int \mathcal{D}\psi \mathcal{D}\Bar{\psi}~ \rme^{-S_{2d}^{\mathrm{fermion}} [\psi, \bar{\psi}]} \Bigl[ \delta_{m_1',m_1}\delta_{m_2',m_2} \notag \\
    &~~~~~+ \int\rmd^2\Vec{x} ~\mathcal{V}_1(\Vec{x}) \delta_{m_1',m_1+1}\delta_{m_2',m_2} + \int\rmd^2\Vec{x} ~\mathcal{V}_1^*(\Vec{x}) \delta_{m_1',m_1-1}\delta_{m_2',m_2} \notag \\
     &~~~~~+ \int\rmd^2\Vec{x} ~\mathcal{V}_2(\Vec{x}) \delta_{m_1',m_1}\delta_{m_2',m_2+1} + \int\rmd^2\Vec{x} ~\mathcal{V}_2^*(\Vec{x}) \delta_{m_1',m_1}\delta_{m_2',m_2-1}
    \Bigr], \label{eq:one_instanton_amplitude}
\end{align}
where the Kronecker delta is understood in $\mathbb{Z}_N$.
Here, let us change the basis with $\{ \ket{k_1,k_2}; k_1,k_2 \in \mathbb{Z}_N \}$:
\begin{align}
    \ket{k_1,k_2} = \frac{1}{N} \sum_{m_1,m_2 \in \mathbb{Z}_N} \rme^{\frac{2 \pi \im k_1 m_1}{N} + \frac{2 \pi \im k_2 m_2}{N}} \ket{m_1,m_2} \label{eq:semiclassical_vacua}
\end{align}
On this basis, the 1-fractional-instanton amplitude can be rewritten as,
\begin{align}
    &\braket{k_1',k_2'|\rme^{-T \hat{H}}|k_1,k_2} \sim  \delta_{k_1',k_1}\delta_{k_2',k_2} \int \mathcal{D}\psi \mathcal{D}\Bar{\psi}~\rme^{-S_{2d}^{\mathrm{fermion}} [\psi, \bar{\psi}]} \Bigl[ 1+  \notag \\
    &~~~~~+ \int \rmd^2\Vec{x} ~(\rme^{-\frac{2\pi\im k_1}{N}} \mathcal{V}_1(\Vec{x})  + \rme^{\frac{2\pi\im k_1}{N}}\mathcal{V}_1^*(\Vec{x}) + \rme^{-\frac{2\pi \im k_2}{N}}\mathcal{V}_2(\Vec{x}) + \rme^{\frac{2\pi \im k_2}{N}} \mathcal{V}_2^*(\Vec{x}) )
    \Bigr],
\end{align}
then, by summing up multiple-fractional-instanton amplitudes, we would have
\begin{align}
    &\braket{k_1',k_2'|\rme^{-T \hat{H}}|k_1,k_2} \sim  \delta_{k_1',k_1}\delta_{k_2',k_2} \int \mathcal{D}\psi \mathcal{D}\Bar{\psi} ~\rme^{-S_{2d}^{\mathrm{fermion}} [\psi, \bar{\psi}]}   \notag \\
    &~~~~~\times \rme^{ \int \rmd^2\Vec{x} ~(\rme^{-\frac{2\pi \im k_1}{N}} \mathcal{V}_1(\Vec{x})  + \rme^{\frac{2\pi \im k_1}{N}}\mathcal{V}_1^*(\Vec{x}) + \rme^{-\frac{2\pi \im k_2}{N}}\mathcal{V}_2(\Vec{x}) + \rme^{\frac{2\pi \im k_2}{N}} \mathcal{V}_2^*(\Vec{x}) )},
\end{align}
Therefore, the states $\{ \ket{k_1,k_2}; k_1, k_2 \in \mathbb{Z}_N \}$ (\ref{eq:semiclassical_vacua}) correspond to the semiclassical vacua appearing in (\ref{eq:dilute_gas_paritition_fct}).
This explains how the $N^2$ semiclassical vacua are constructed in the effective quantum mechanics with $\mathbb{R} \times S^1_A \times (S^1_B \times S^1_C)$ compactification.

In the $SU(N)_1 \times SU(N)_2$ pure Yang-Mills theory, $\ket{k_1 = 0, k_2 = 0}$ is the vacuum for $-\pi \leq \theta_1 \leq +\pi$ and $-\pi \leq \theta_2 \leq +\pi$, and $\ket{k_1 = 1, k_2 = 0}$ is the vacuum for $\pi \leq \theta_1 \leq +3\pi$ and $-\pi \leq \theta_2 \leq +\pi$, etc.
In the dual superconductor picture, the state $\ket{k_1 = 0, k_2 = 0}$ is the monopole-condensed vacuum for $SU(N)_1$ and $SU(N)_2$, the state $\ket{k_1 = 1, k_2 = 0}$ means the electric-charge-$(-1)$ dyon-condensation for $SU(N)_1$ and the monopole-condensation for $SU(N)_2$, and so on.

\subsection{Massless case}
\label{sec:semiclassics_massless}

In this section, employing the above dilute gas approximation (\ref{eq:dilute_gas_paritition_fct}), we derive the phase diagram of the massless QCD(BF).
For simplicity, let us assume $g_1 = g_2$ (common dynamical scale in $SU(N)_1\times SU(N)_2$), which yields $K^{(1)} = K^{(2)} =: K$ and $S^{(1)}_\mathrm{I} = S^{(2)}_\mathrm{I} =: S_\mathrm{I}$.

The dilute gas picture indicates $N^2$ classical vacua with dynamical fermion, and the classical degenerate $N^2$ vacua split due to the fluctuation of the dynamical fermion.
In the chiral limit $m = 0$, the effective 2d fermion mass is given by center vortex contribution $M_{k_1,k_2}$ in (\ref{eq:effective_mass_massless}).
To identify ground states, we shall evaluate the vacuum energy of fermion living in each classical vacuum.
The partition function associated with classical vacuum $ \ket{k_1,k_2} $ is,
\begin{align}
    Z^{(k_1,k_2)} &:= \int \mathcal{D}\psi \mathcal{D}\Bar{\psi} ~\rme^{-S_{2d}^{(k_1,k_2)}[\psi, \bar{\psi}]} \notag \\
    &= \int \mathcal{D}\psi \mathcal{D}\Bar{\psi} ~\rme^{-  \begin{pmatrix}
\bar{\psi}_R & \Bar{\psi}_L \\
\end{pmatrix}
\begin{pmatrix}
M_{k_1,k_2}^* &   (\partial_1 -\im \partial_2)  \\
(\partial_1 + \im \partial_2)  & M_{k_1,k_2} \\
\end{pmatrix}
\begin{pmatrix}
\psi_L \\
\psi_R 
\end{pmatrix}
} \notag \\
&= \left[ \operatorname{Det} \left( -\partial^2_{2d} + |M_{k_1,k_2}|^2 \right) \right]^2 =: \rme^{- V \mathcal{E}_{\mathrm{vac}}^{(k_1,k_2)} } \notag \\
\end{align}
where $ \mathcal{E}_{\mathrm{vac}}^{(k_1,k_2)} $ is the vacuum energy density of the $(k_1,k_2)$ sector, given by
\begin{align}
    \mathcal{E}_{\mathrm{vac}}^{(k_1,k_2)} &= -2 \int \frac{\mathrm{d}^2 p}{(2\pi)^2} \log (p^2 + |M_{k_1,k_2}|^2) \notag \\
    &= -2 \left[ \frac{\Lambda^2_\mathrm{UV}}{4 \pi} \left( \log \Lambda^2_\mathrm{UV} - 1 \right) + \frac{1}{4 \pi} |M_{k_1,k_2}|^2 \left( \log \left( \frac{\Lambda^2_\mathrm{UV}}{|M_{k_1,k_2}|^2} \right) + 1 \right) + O(\Lambda^{-2}_\mathrm{UV})\right], \label{eq:vacuum_energy_massless}
\end{align}
with an ultraviolet cutoff $\Lambda_\mathrm{UV}$.
Above the scale $\sim 1/(NL)$, the 4d structure becomes relevant, e.g., higher Kaluza-Klein modes start to contribute.
Hence, the cutoff of the 2d effective theory should be set as this scale: $\Lambda_\mathrm{UV} \sim 1/(NL)$.
Note that $|M_{k_1,k_2}| \sim \rme^{-S_\mathrm{I}/N}/L\sim (NL\Lambda_\mathrm{UV})^3/L$ becomes much smaller than $\Lambda_\mathrm{UV}$ when $NL\Lambda_\mathrm{UV}\ll 1$.

In the expression of the vacuum energy (\ref{eq:vacuum_energy_massless}), the first term $ \frac{\Lambda^2_\mathrm{UV}}{4 \pi} \left( \log \Lambda^2_\mathrm{UV} - 1 \right)$ is independent of $(k_1,k_2)$, which can be absorbed to an overall factor of the total partition function.
The second term $ \frac{1}{4 \pi} |M_{k_1,k_2}|^2 \left( \log \left( \frac{\Lambda^2_\mathrm{UV}}{|M_{k_1,k_2}|^2} \right) + 1 \right)$ splits the $N^2$ degenerate classical vacua: $\ket{k_1,k_2}$ with largest $|M_{k_1,k_2}|$ is the ground state.

Based on this observation, we can draw the phase diagram.
Immediately, we can find the following qualitative features:
\begin{itemize}
    \item The energy levels depend only on $|M_{k_1,k_2}|$, the vacuum structure is invariant under the shift $(\theta_1, \theta_2) \rightarrow (\theta_1 + \alpha, \theta_2 + \alpha)$, as expected.
    \item Similarly, the $N$ pair $(k_1,k_2), (k_1 + 1, k_2 + 1), \cdots, (k_1 + N-1, k_2 + N-1)$ have the same energy level:
\begin{align}
    \mathcal{E}_{\mathrm{vac}}^{(k_1,k_2)} = \mathcal{E}_{\mathrm{vac}}^{(k_1+1,k_2+1)} = \cdots = \mathcal{E}_{\mathrm{vac}}^{ (k_1 + N-1, k_2 + N-1)},
\end{align}
    leading to (at least) $N$ degenerate vacua. In other words, the vacuum energy of $(k_1,k_2)$ sector depends only on $k_1 - k_2$.
\end{itemize} 

\begin{figure}[t]
\centering
\includegraphics[scale=0.6]{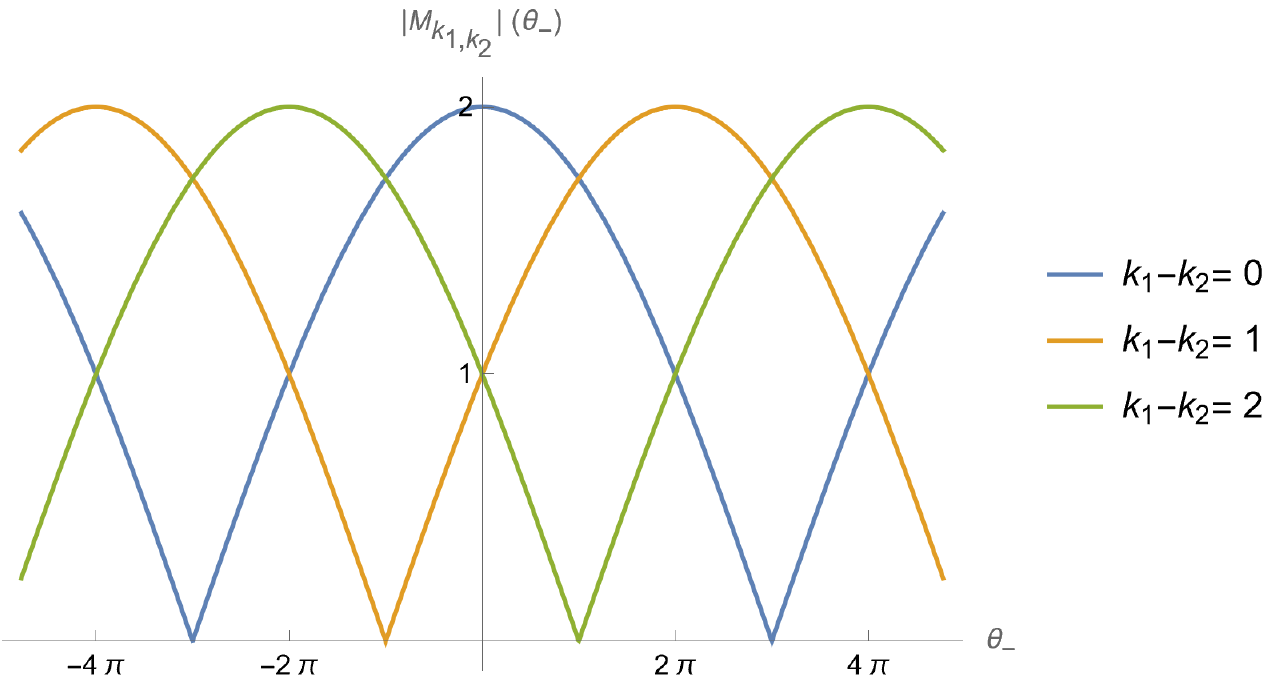}
\caption{
Mass function $|M_{k_1,k_2}|(\theta_-)$ at $N=3$ with normalization $K \rme^{- S_\mathrm{I}/N} = 1$.
Each line stands for $3$-fold ($N$-fold) degenerate states; for example, the $k_1 - k_2 = 0$ line corresponds to $(k_1,k_2) = (0,0), (1,1), (2,2)$ states. This plot shows that the $k_1 - k_2 = 0$ states are ground states for $-\pi < \theta_- < +\pi $, the $k_1 - k_2 = 1$ states are ground states for $\pi < \theta_- < +3\pi $, etc.
The ground states are $2N$-fold vacua at the phase boundary.
For example, at $\theta_- = \pi$, the 6 states with $k_1 - k_2 = 0$ or $k_1 - k_2 = 1$ are the ground states.
}
\label{fig:N3_massfunction}
\end{figure}

Then, we can easily find which sector gives the minimum: 
\begin{align}
    |M_{k_1,k_2}|(\theta_-) &= K \rme^{- S_\mathrm{I}/N} |1 + \rme^{\im (\theta_- - 2 \pi (k_1 - k_2))/N}| \notag \\
    &= 2  K \rme^{- S_\mathrm{I}/N} \left| \cos \left( \frac{\theta_- - 2 \pi (k_1 - k_2)}{2N} \right) \right|.
\end{align}
with $\theta_- = \theta_1 - \theta_2$. Therefore, for $-\pi + 2 \pi n< \theta_- < +\pi + 2 \pi n$, the $N$ degenerate states with $(k_1 - k_2) = n ~(\operatorname{mod}N)$ gives the lowest energy.
This feature is illustrated in Fig.~\ref{fig:N3_massfunction}, which plots the mass function $|M_{k_1,k_2}|(\theta_-)$ at $N=3$.

Hence, the semiclassical picture indeed explains the conjectured phase diagram of the massless QCD(BF), Fig.~\ref{fig:conjectured_massless}, including $N$-fold degeneracy (and $2N$-fold degeneracy on the phase boundary). 
In particular, we would like here to emphasize that the $\mathbb{Z}_2$ exchange symmetry is unbroken at $\theta_-=0$ since the selected ground states satisfy $k_1=k_2$, and this is consistent with the result of the different semiclassical analysis of QCD(BF) on $\mathbb{R}^3\times S^1$ with the center-stabilizing potential~\cite{Shifman:2008ja}. 

Before moving to the massive case, let us comment on cases with $g_1 \neq g_2$, or different dynamical scales for $SU(N)_1 \times SU(N)_2$.
This only affects the evaluation of the mass function, that is,
\begin{align}
    |M_{k_1,k_2}|(\theta_-) &= K^{(2)} \rme^{- S_\mathrm{I}^{(2)}/N} |1 + R \rme^{\im (\theta_- - 2 \pi (k_1 - k_2))/N}| , \notag \\
    R &:= \frac{K^{(1)} \rme^{- S_\mathrm{I}^{(1)}/N}}{K^{(2)} \rme^{- S_\mathrm{I}^{(2)}/N}} > 0, 
\end{align}
which gives the same lowest-energy states as $g^{(1)} = g^{(2)}$.
Hence, the phase diagram is coherently given by Fig.~\ref{fig:conjectured_massless}, irrespective of different dynamical scales.

\subsection{Mass perturbation}
\label{sec:semiclassics_massive}

Next, we include the mass perturbation $m$ to the bifundamental fermion, and we assume that the bare mass is much smaller than the Kaluza-Klein scale, 
\begin{align}
    m \ll 1/NL.
\end{align}
Within this mass hierarchy, we can apply the semiclassical results described in Section~\ref{sec:methodology}.
Adding the mass term $m \bar{\psi} \psi = m (\bar{\psi}_L\psi_R + \bar{\psi}_R\psi_L)$, the action for $(k_1,k_2)$ sector (\ref{eq:sector_action}) is given by
\begin{align}
    S_{2d}^{(k_1,k_2)}[\psi, \bar{\psi}] &= S_{2d}^{\mathrm{fermion}}[\psi, \bar{\psi}] +  \int\rmd^2\Vec{x} ~\Bigl[  M_{k_1,k_2} (\bar{\psi}_L \psi_R)(\Vec{x}) + M_{k_1,k_2}^* (\bar{\psi}_R \psi_L)(\Vec{x})  \Bigr] \notag \\
    &= \int \diff^2\vec{x}
\begin{pmatrix}    
    \bar{\psi}_R & \Bar{\psi}_L \\
\end{pmatrix}
\begin{pmatrix}
M_{k_1,k_2}^* +m & (\partial_1 -\im \partial_2)  \\
(\partial_1 + \im\partial_2)  & M_{k_1,k_2} +m\\
\end{pmatrix}
\begin{pmatrix}
\psi_L \\
\psi_R 
\end{pmatrix}.
\end{align}
We can basically repeat the argument in the previous section applies to obtain the phase diagram by taking into account the modification of the effective mass $M_{k_1,k_2}$:
\begin{align}
    M_{k_1,k_2} \rightarrow \Tilde{M}_{k_1,k_2} = M_{k_1,k_2} + m
\end{align}
Our task is thus to find $(k_1,k_2)$ with maximal $|\Tilde{M}_{k_1,k_2}|$.
Unlike the massless case, the phase of $M_{k_1,k_2}$ becomes important because of the mass $m > 0$, which singles out one vacuum except for the transition line.

For example, without current fermion mass $m = 0$, the $(0,0), \cdots, (N-1,N-1)$ vacua form the $N$ degenerate ground states at $(\theta_1, \theta_2) = 0$: same absolute values $|M_{0,0}| = |M_{1,1}| = \cdots = |\Tilde{M}_{N-1,N-1}|$ with different phases $M_{k,k} = \rme^{- \frac{2 \pi \im k}{N}}|M_{k,k}|$.
In the presence of non-zero mass $m > 0$, the vacuum energy depends on $|\Tilde{M}_{k_1,k_2}| = |M_{k_1,k_2} + m|$ instead of $|M_{k_1,k_2}|$.
Hence, among the classical vacua $\{ \ket{k,k} \}_{k = 0, \cdots, N-1}$, the vacuum $\ket{0,0}$ with $M_{k,k} > 0$ gives the largest $|\Tilde{M}_{k_1,k_2}|$, which is indeed the ground state.
By increasing $\theta_+ = \theta_1 + \theta_2$ with fixed $\theta_- =  \theta_1 - \theta_2 = 0$, the phase of $M_{k,k}$ becomes $M_{k,k} = \rme^{\im \frac{\theta_+/2 - 2 \pi k}{N}}|M_{k,k}|$.
Therefore, the $(0,0)$ vacuum is the ground state for $-2\pi < \theta_+ < 2 \pi$, the $(1,1)$ vacuum is the ground state for $2\pi < \theta_+ < 4 \pi$, etc.

\begin{figure}[t]
\centering
\begin{minipage}{.45\textwidth}
\subfloat[small-mass case $m \ll K\rme^{-S_\mathrm{I}/N}$]{
\includegraphics[scale=0.45]{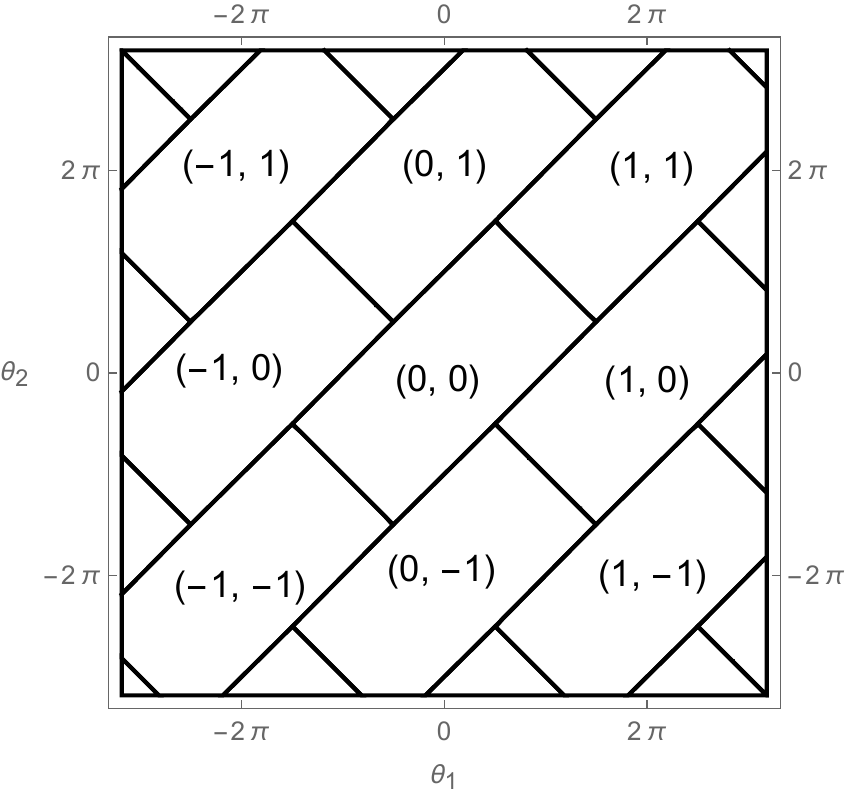}
\label{fig:near_massless_phase_diagram}
}\end{minipage}\quad
\begin{minipage}{.45\textwidth}
\subfloat[``Large-mass'' case $K\rme^{-S_\mathrm{I}/N} \ll m \ll 1/NL$]{\includegraphics[scale=0.45]{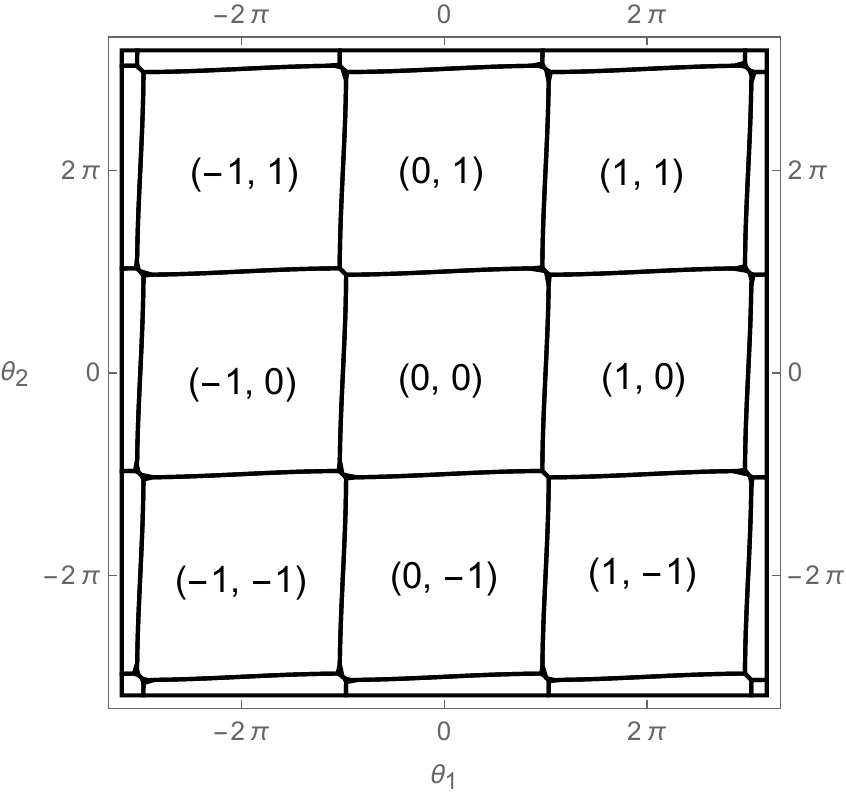}
\label{fig:large_mass_phase_diagram}}
\end{minipage}
\caption{Phase diagrams obtained by the semiclassics for small mass and large mass limits. 
(a) Phase diagram for small-mass case $m \ll K\rme^{-S_\mathrm{I}/N}$.
The label $(k_1,k_2)$ stands for the $(k_1,k_2) \in \mathbb{Z}_N \times \mathbb{Z}_N$ classical vacuum.
The $N$ degeneracy in the massless case (Fig.~\ref{fig:conjectured_massless}) is resolved by the mass perturbation.
(b) ``Large-mass'' case $K\rme^{-S_\mathrm{I}/N} \ll m \ll 1/NL$. This phase diagram can be understood as a small deformation from that of the pure $SU(N) \times SU(N)$ Yang-Mills theory.}
\label{fig:small_large_mass}
\end{figure}

Based on this observation, let us see small-mass ($m \ll K\rme^{-S_\mathrm{I}/N}$) and large-mass ($K\rme^{-S_\mathrm{I}/N} \ll m \ll 1/NL$) limits. The results are depicted in Fig.~\ref{fig:small_large_mass}.

\begin{itemize}
    \item Small mass

    In this case, the above discussion at $\theta_- = 0$ can be applied for all $\theta_-$.
More explicitly, since
\begin{align}
    |\Tilde{M}_{k_1,k_2}|^2 = |M_{k_1,k_2}|^2 + 2 m \operatorname{Re} M_{k_1,k_2} + m^2, \label{eq:M_tilde_squared}
\end{align}
we deduce that, for states with identical $|M_{k_1,k_2}|$, a state with larger $ \operatorname{Re} M_{k_1,k_2}$ is preferred for small $m$.
We then obtain the phase diagram with small mass perturbation, shown in (Fig.~\ref{fig:near_massless_phase_diagram}).

    \item Large mass

    To deal with this case semiclassically, we assume the mass hierarchy: $K\rme^{-S_\mathrm{I}/N} \ll m \ll 1/NL$.
    In this case, the significance of each term in (\ref{eq:M_tilde_squared}) is reversed:
    First, states with larger $\operatorname{Re} M_{k_1,k_2}$ are favored, and then, the absolute-value term $|M_{k_1,k_2}|^2$ perturbs the energy levels. 
    Since
\begin{align}
    \operatorname{Re} M_{k_1,k_2} = K \rme^{-S_\mathrm{I}/N} \left[ \cos \left( \frac{\theta_1 - 2 \pi k_1}{N} \right) + \cos \left( \frac{\theta_2 - 2 \pi k_2}{N} \right)\right],
\end{align}
    the phase boundary becomes the square-lattice pattern at the leading order as in the case of the $SU(N) \times SU(N)$ pure Yang-Mills theory.
    The subleading contribution becomes important at the 4-fold degenerate point such as $\theta_1 = \theta_2 = \pi$, where the four states $(0,0)$, $(1,0)$, $(0,1)$, and $(1,1)$ has the same $\operatorname{Re} M_{k_1,k_2}$. The absolute-value term $|M_{k_1,k_2}|^2$ partially splits the degenerate states, and $(0,0)$ and $(1,1)$ states are favored.
    As a result, we obtain the phase diagram given in Fig.~\ref{fig:large_mass_phase_diagram}. 
\end{itemize}

These large- and small-mass limits show that the phase diagram indeed has the same topology as the conjectured one, Fig.~\ref{fig:conjectured_massive}. 
We note that this semiclassical analysis is reliable only when $NL\Lambda\ll 1$, but we are tempted to believe that the $4$-dimensional strongly-coupled theory shows the same qualitative behavior under the adiabatic continuity.

\subsection{Phase diagrams at several parameter setups}
\label{sec:intermediate_m}

\begin{figure}[t] 
  \begin{minipage}[b]{0.5\linewidth}
    \centering \subfloat[$m = 0.3$]{
\includegraphics[scale=0.4]{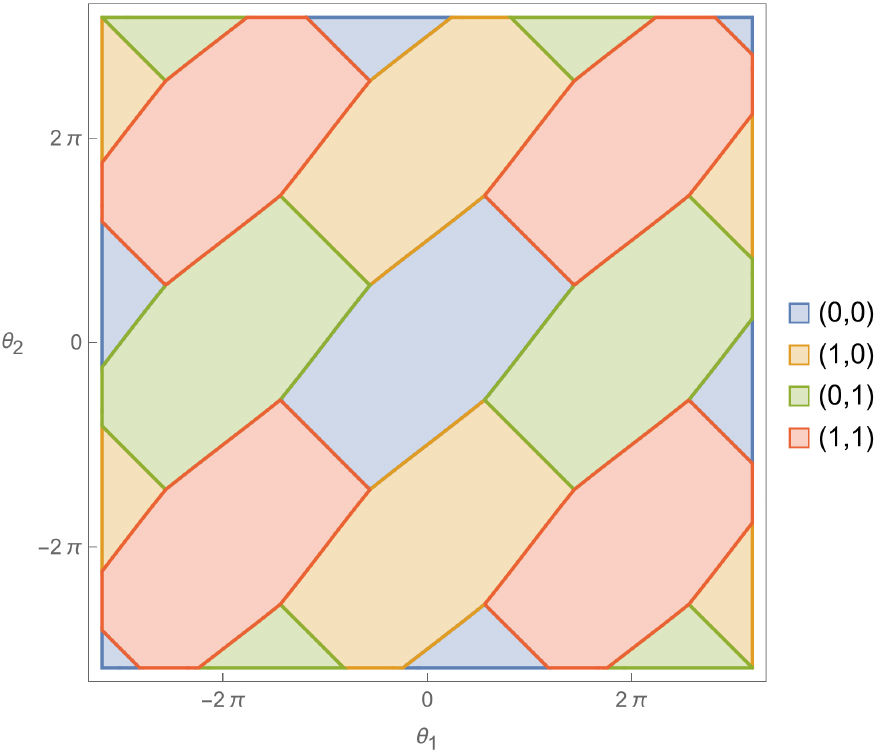}
}
    \vspace{4ex}
  \end{minipage}
  \begin{minipage}[b]{0.5\linewidth}
    \centering
    \subfloat[$m = 1$]{
\includegraphics[scale=0.4]{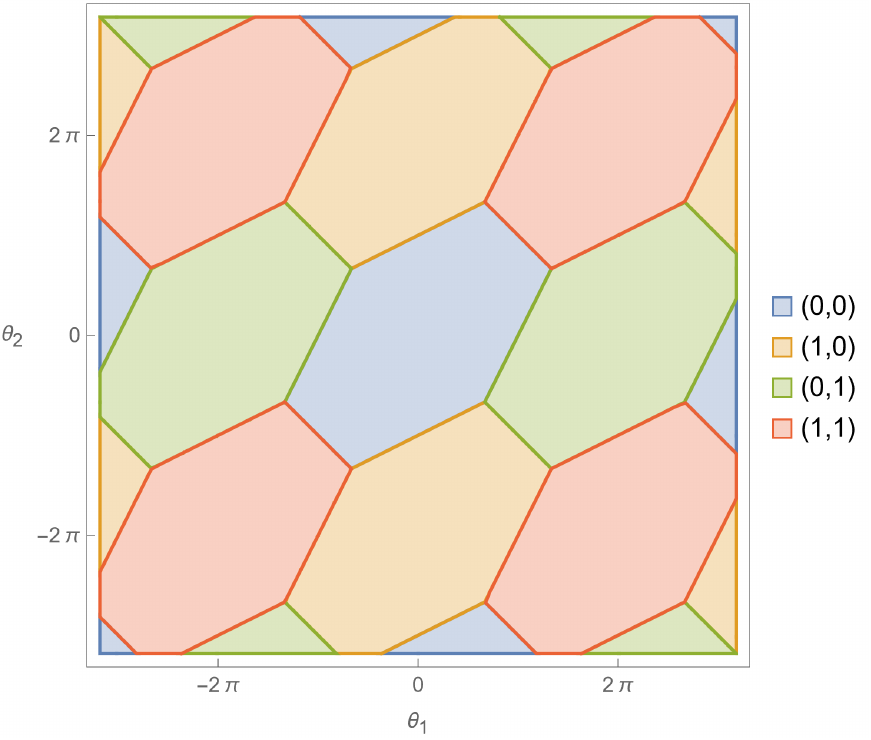}
}
    \vspace{4ex}
  \end{minipage} 
  \begin{minipage}[b]{0.5\linewidth}
    \centering
    \subfloat[$m = 3$]{
\includegraphics[scale=0.4]{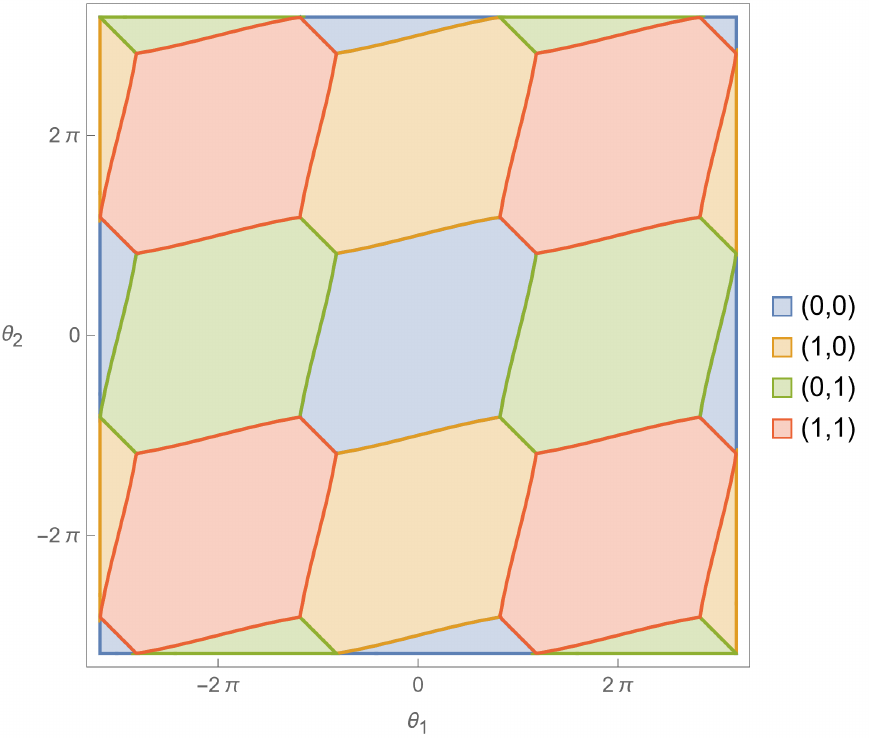}
\label{fig:app_m3}
}
    \vspace{4ex}
  \end{minipage}
  \begin{minipage}[b]{0.5\linewidth}
    \centering
    \subfloat[$m = 5$]{
\includegraphics[scale=0.4]{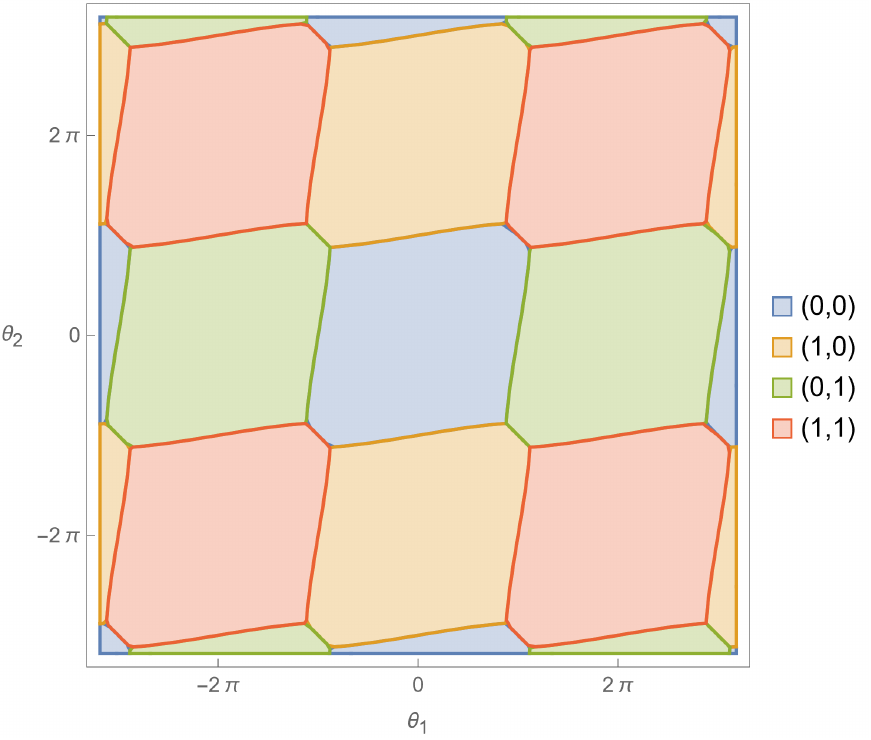}
}
    \vspace{4ex}
  \end{minipage} 
  \caption{Phase diagrams at $m = 0.3,~1,~3,~5$ with mass unit $K\rme^{-S_\mathrm{I}/N} = 1$ for $N=2$ and symmetric dynamical scales $\Lambda_1 =\Lambda_2$.
  The blue, yellow, green, and red regions represent $(k_1,k_2) = (0,0)$, $(1,0)$, $(0,1)$, and $(1,1)$ phases, respectively.
  (a): phase diagram at $m = 0.3$. The shape of the phase boundary is still similar to that of the small-mass limit, Fig.~\ref{fig:near_massless_phase_diagram}. (b) and (c): phase diagrams at $m = 1$ and $m=3$. The phase boundary is gradually converging to the square lattice. (d):phase diagram at $m = 5$. Now, the shape of the phase boundary is still similar to that of the large-mass limit, Fig.~\ref{fig:large_mass_phase_diagram}. }
  \label{fig:app-intermediate-m} 
\end{figure}

In the rest of this section, we show phase diagrams with several concrete parameter setups, including intermediate fermion mass $m$ and scale hierarchy $\Lambda_1 \neq \Lambda_2$.
Let us first present phase diagrams at intermediate $m$.
For concreteness, we consider $N = 2$ cases with normalization $K\rme^{-S_\mathrm{I}/N} = 1$ for numerical computations.
In these cases, we have 4 semiclassical vacua: $(0,0)$, $(1,0)$, $(0,1)$, and $(1,1)$, and Fig.~\ref{fig:app-intermediate-m} depicts which semiclassical vacua have the lowest energy.
These phase diagrams illustrate how the small-mass phase diagram (Fig.~\ref{fig:near_massless_phase_diagram}) is continuously deformed into the large-mass limit (Fig.~\ref{fig:large_mass_phase_diagram}) as $m$ increases.

\begin{figure}[t] 
  \begin{minipage}[b]{0.5\linewidth}
    \centering \subfloat[$K^{(2)} \rme^{-S^{(2)}_\mathrm{I} / N} = 1$]{
\includegraphics[scale=0.4]{Figures/m30.pdf}
}
    \vspace{4ex}
  \end{minipage}
  \begin{minipage}[b]{0.5\linewidth}
    \centering
    \subfloat[$K^{(2)} \rme^{-S^{(2)}_\mathrm{I} / N} = 10$]{
\includegraphics[scale=0.4]{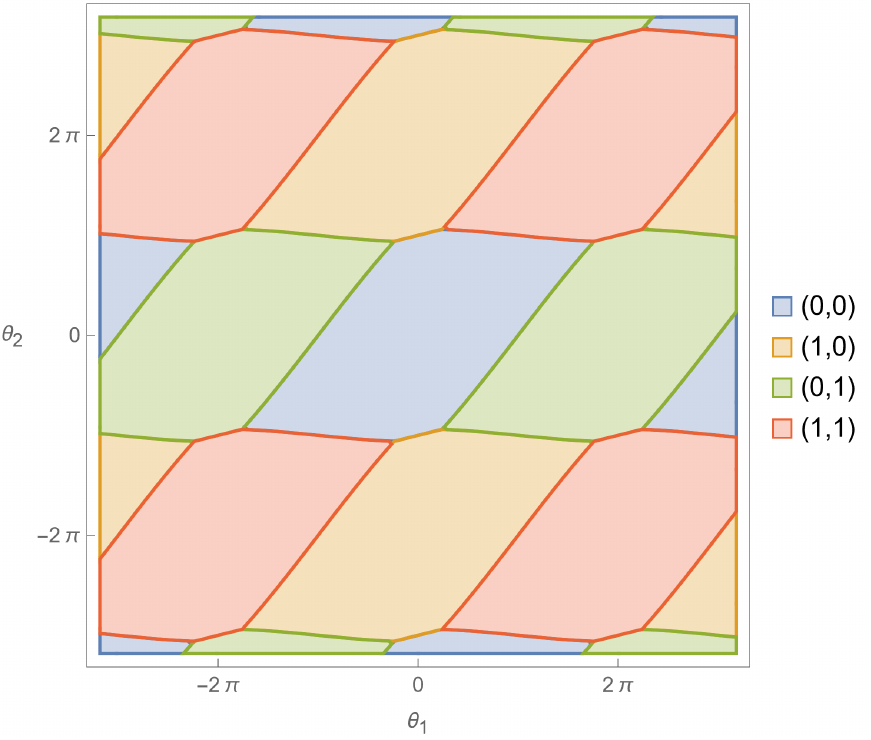}
}
    \vspace{4ex}
  \end{minipage} 
  \begin{minipage}[b]{0.5\linewidth}
    \centering
    \subfloat[$K^{(2)} \rme^{-S^{(2)}_\mathrm{I} / N} = 100$]{
\includegraphics[scale=0.4]{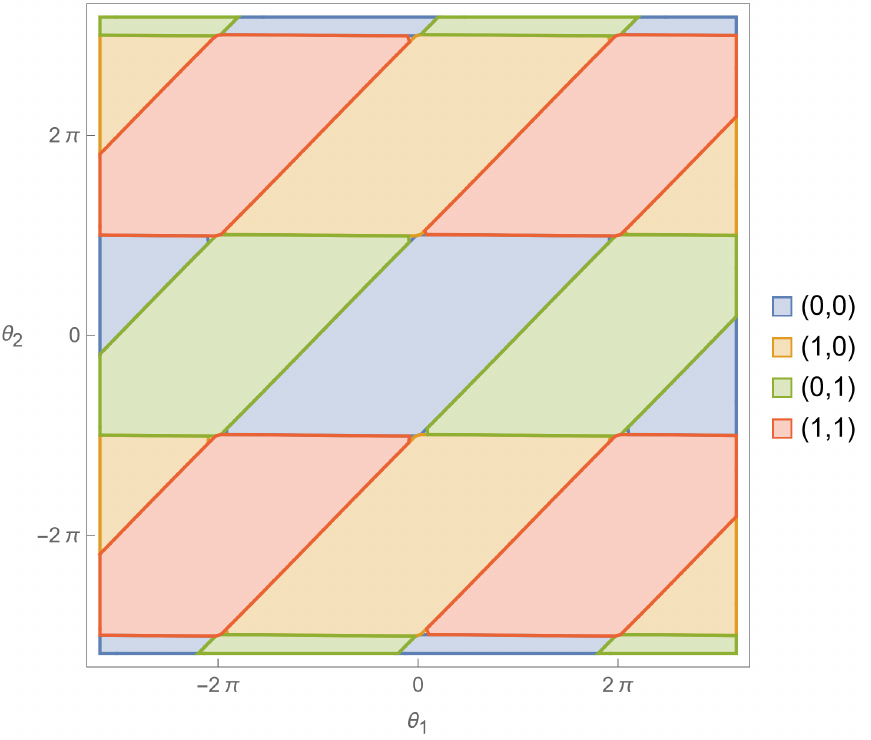}
}
    \vspace{4ex}
  \end{minipage}
  \begin{minipage}[b]{0.5\linewidth}
    \centering
    \subfloat[$K^{(2)} \rme^{-S^{(2)}_\mathrm{I} / N} \rightarrow \infty$ (or $\Lambda_1, m \ll \Lambda_2$)]{
\includegraphics[scale=0.4]{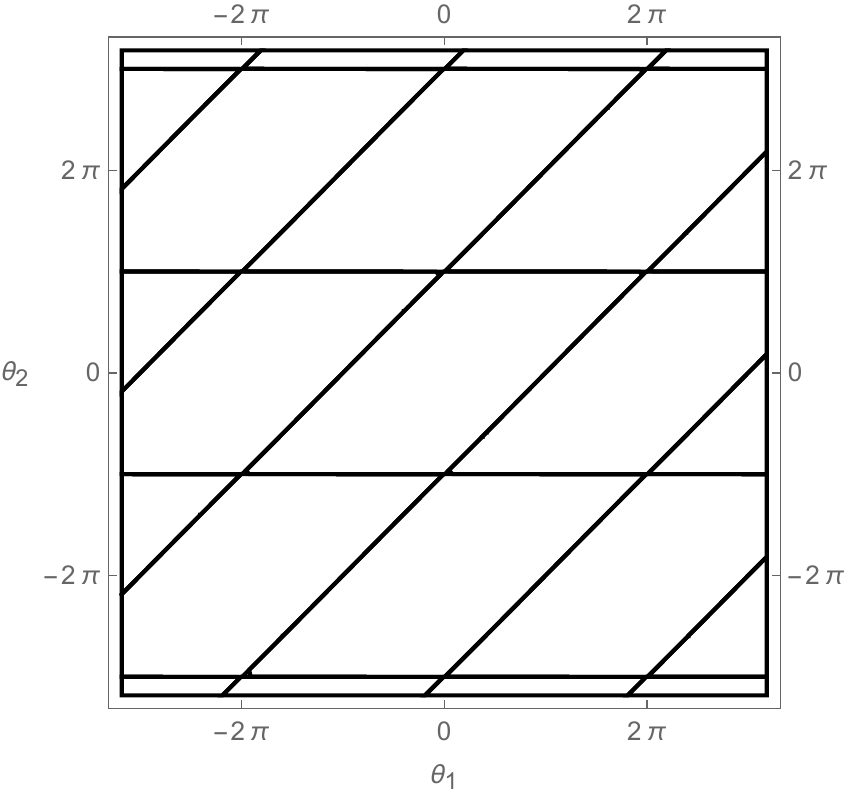}
\label{fig:hierarchical_limit}
}
    \vspace{4ex}
  \end{minipage} 
  \caption{
  Phase diagrams at $m = 3$ for $N=2$ with several values of $K^{(2)} \rme^{-S^{(2)}_\mathrm{I} / N}$ (in the unit of $K^{(1)} \rme^{-S_\mathrm{I}^{(1)}/N} = 1$) and the hierarchical limit $K^{(2)} \rme^{-S^{(2)}_\mathrm{I} / N} \rightarrow + \infty$. The phase diagrams are colored in the same way as before, Fig.~\ref{fig:app-intermediate-m}.
  (a) symmetric case $K^{(2)} \rme^{-S^{(2)}_\mathrm{I} / N} = 1$, same as Fig.~\ref{fig:app_m3},  (b) $K^{(2)} \rme^{-S^{(2)}_\mathrm{I} / N} = 10$,
  (c) $K^{(2)} \rme^{-S^{(2)}_\mathrm{I} / N} = 100$,
  (d) phase boundary in the hierarchical limit $K^{(2)} \rme^{-S^{(2)}_\mathrm{I} / N} \rightarrow + \infty$, explained in the main text.}
  \label{fig:app-hierarchcal} 
\end{figure}

Next, we discuss phase diagrams with different dynamical scales $\Lambda_1 \neq \Lambda_2$, or different coupling $g_1 \neq g_2$.
Here, we choose mass unit $K^{(1)} \rme^{-S_\mathrm{I}^{(1)}/N} = 1$ and vary $K^{(2)} \rme^{-S_\mathrm{I}^{(2)}/N}$.
Increasing $K^{(2)} \rme^{-S_\mathrm{I}^{(2)}/N}$ corresponds to increasing $\Lambda_2$, compared to $\Lambda_1$.
To illustrate a typical trend, we present Fig.~\ref{fig:app-hierarchcal}, which depicts phase diagrams at $m = 3$ (in the above mass unit) with several values of $K^{(2)} \rme^{-S^{(2)}_\mathrm{I} / N} = 1,~10,~100$. 

The last figure, Fig.~\ref{fig:hierarchical_limit}, shows the phase diagram at the hierarchical limit $\Lambda_1, m \ll \Lambda_2$, which can be understood as follows.
As before, we shall find $(k_1,k_2)$ with maximal $|\Tilde{M}_{k_1,k_2}| = |M_{k_1,k_2} +m|$.
In the limit $K^{(2)} \rme^{-S^{(2)}_\mathrm{I} / N} \rightarrow \infty$, the dominant contribution of $|\Tilde{M}_{k_1,k_2}|^2$ is given by
\begin{align}
    |\Tilde{M}_{k_1,k_2}|^2 &= |K^{(2)} \rme^{-S^{(2)}_\mathrm{I} / N}|^2 \notag \\
    &+ 2 K^{(2)} \rme^{-S^{(2)}_\mathrm{I} / N} \left[ m \cos \left( \frac{\theta_2 - 2 \pi k_2}{N}\right) + K^{(1)} \rme^{-S^{(1)}_\mathrm{I} / N} \cos \left( \frac{\theta_1 - \theta_2 - 2 \pi (k_1-k_2)}{N}\right) \right] \notag \\
    &+ O(1),
\end{align}
Since the first term, $|K^{(2)} \rme^{-S^{(2)}_\mathrm{I} / N}|^2$, is just an overall constant, what determines the ground state $(k_1,k_2)$ is the second term, $O(K^{(2)} \rme^{-S^{(2)}_\mathrm{I} / N})$.
This $(k_1,k_2)$ dependence yields the phase diagram at the hierarchical limit $\Lambda_2 \rightarrow \infty$, Fig.~\ref{fig:hierarchical_limit}.
This phase boundary for the hierarchical dynamical scales is consistent with the prediction of the four-dimensional chiral effective Lagrangian (see Section~6 of Ref.~\cite{Karasik:2019bxn}).

\section{Discussions}
\label{sec:discussion}
\subsection{On domain walls at the phase boundaries}

In Section~\ref{Sec:semiclassics}, we obtained the phase diagram of the QCD(BF) in the semiclassically-calculable confinement regime and found the rich phase structure. 
As a function of $\theta_1$ and $\theta_2$, $N^2$ vacua labeled by $k_1, k_2$ are exchanged by the 1st-order quantum phase transitions. 
Then, we consider the situation where the domain wall connects different vacua, and the domain wall is a dynamical object with an interesting structure in the 4-dimensional theory (see, e.g., Refs.~\cite{Anber:2015kea, Sulejmanpasic:2016uwq, Komargodski:2017smk, Gaiotto:2017tne}).
Here, let us comment on domain walls connecting these vacua in the semiclassical description.

Let us compare the states with $(k_1,k_2)$ and $(k'_1,k'_2)=(k_1+\Delta k_1, k_2+\Delta k_2)$. 
The anomaly polynomial suggests that these two states have different contact terms for the background gauge fields,
\begin{equation}
    \Delta S_{\mathrm{top}}=\frac{N(\Delta k_1+\Delta k_2)}{4\pi}\int_{M_4} B^2-\frac{N \Delta k_2}{2\pi}\int_{M_4} B\wedge (\diff A-B_A). 
\end{equation}
In the $2$d reduction $M_4=M_2\times T^2$ with the minimal 't~Hooft flux $(n_{34}=1)$, this topological action becomes 
\begin{equation}
    \Delta S_{\mathrm{top}}=(\Delta k_1+\Delta k_2)\int_{M_2} B - \Delta k_2\int_{M_2} (\diff A-B_A).  
    \label{eq:2dDomainWall_Anomaly}
\end{equation}
The domain-wall theory connecting $|k_1,k_2\rangle$ and $|k'_1,k'_2\rangle$ must cancel the anomaly-inflow from this effective $2$d topological action.  

In the $2$d effective theory~\eqref{eq:effective_mass_massless}, these vacuum labels $k_1,k_2$ appear as the discrete labels instead of the expectation values of some local operators in the effective description. 
In general, such labels correspond to the expectation values of the heavy fields that are already integrated out, and thus, unfortunately, we cannot consider the dynamical domain wall connecting different $k_1,k_2$ within our effective theory. 
Instead, we need to treat the domain wall as the external operator, and this role is taken by the Wilson loops in our case. 

Let us consider the Wilson loop operators, $W_1$ and $W_2$, for $2$d $SU(N)_{1,2}$ gauge fields, and then we introduce the loop operator $W_1^{\Delta k_1} W_2^{\Delta k_2}(C)$. 
When the vacuum outside the loop is given by $|k_1,k_2\rangle$, the vacuum inside the loop becomes $|k'_1, k'_2\rangle$, since these labels $k_1, k_2$ basically measure the $\mathbb{Z}_N$ $1$-form symmetry charges of $SU(N)_{1,2}$, respectively. 
As the bifundamental matter only maintains the diagonal center, this Wilson loop has the charge $\Delta k_1+\Delta k_2$ of the $\mathbb{Z}_N^{[1]}$ symmetry, and this correctly matches the anomaly inflow from the first term of the topological action~\eqref{eq:2dDomainWall_Anomaly}. 
Recalling \eqref{eq:couplingsBackgroundFields}, we can readily see that it also matches the second term of \eqref{eq:2dDomainWall_Anomaly}. 
We note that the selection rule of $2$d $1$-form symmetry is so strong that the Hilbert space is completely decomposed by its charge, which is called the decomposition or the universe selection rule~\cite{Pantev:2005zs, Hellerman:2006zs, Hellerman:2010fv, Cherman:2019hbq, Tanizaki:2019rbk, Komargodski:2020mxz}. 
Therefore, it is the anomaly that requires the fact that the domain wall becomes non-dynamical when $\Delta k_1+\Delta k_2\not =0$, and this cannot be circumvented as long as we consider the $T^2$ compactification with the 't~Hooft flux. 

Let us next consider the case when $\Delta k_1+\Delta k_2=0$. 
Then, the first term of \eqref{eq:2dDomainWall_Anomaly} vanishes and thus the $2$d theory would be able to support the dynamical domain walls connecting $(k_1, k_2)$ and $(k'_1,k'_2)=(k_1+\Delta k_1, k_2-\Delta k_1)$. 
However, our $2$d effective theory does not contain the local field that can change the value of the semiclassical vacuum labels $k_1, k_2$, and we have to introduce the external Wilson loop $(W_1 W_2^{-1})^{\Delta k_1}$ to achieve it. 
We would like to notice that the bifundamental fermion $\Psi$ has the same gauge charge with the bifundamental loop, $W_1 W_2^{-1}$, and thus $(W_1 W_2^{-1})^{\Delta k_1}$  can be screened by (or endable at) $\Psi^{\Delta k_1}$. 
This consideration suggests that we need to reinstate the higher Kaluza-Klein modes of $\Psi$ to create the dynamical domain wall. 
Since such a particle has the mass gap $O(1/NL)$, the dynamical domain wall of the $2$d theory must also have the energy at least of the order of $1/NL$, which is beyond the scope of the zero-mode effective theory~\eqref{eq:effective_mass_massless}. 
We note that this heavy dynamical domain wall has the fractional $U(1)_V/\mathbb{Z}_N$ charge $\Delta k_1/N$, and it also correctly cancels the anomaly inflow from the second term of \eqref{eq:2dDomainWall_Anomaly}.

\subsection{On large-\texorpdfstring{$N$}{N} orbifold equivalence}

One of the interesting features of QCD(BF) is that it is a candidate for the nonperturbative orbifold equivalence between supersymmetric and non-supersymmetric gauge theories. 
The orbifold equivalence roughly states that the $\mathcal{N} = 1$ $SU(2N)$ super Yang-Mills theory (or $\mathcal{N} = 1$ SYM) and $SU(N) \times SU(N)$ QCD(BF) with $g_1=g_2$ and $\theta_1=\theta_2$ are equivalent about ``neutral-sector'' observables in the large-$N$ limit. 
This large-$N$ equivalence is proven at all orders of the perturbative expansion~\cite{Kachru:1998ys, Bershadsky:1998cb}, and the necessary and sufficient condition for its nonperturbative validity is established by Kovtun, \"{U}nsal and Yaffe~\cite{Kovtun:2003hr,Kovtun:2004bz,Kovtun:2005kh}: 
The large-$N$ nonperturbative equivalence holds if and only if the vacuum of the parent theory ($\mathcal{N} = 1$ $SU(2N)$ SYM) belongs to the orbifold-projection-symmetry invariant sector and that of the daughter theory ($SU(N) \times SU(N)$ QCD(BF)) belong to the quiver-permutation invariant sector.
Since the orbifold projection just uses the fermion parity, the parent side obviously satisfies the condition as long as we believe that the SYM vacuum is Lorentz-invariant. 
Thus, the question of its validity is translated into the question if the QCD(BF) vacuum maintains the $\mathbb{Z}_2$ exchange symmetry.

\subsubsection{Validity of the orbifold equivalence on the bulk physics}
We note that QCD(BF) on $\mathbb{R}^4$ is strongly coupled and thus answering the above question requires to solve the strongly-coupled nonperturbative dynamics. 
Although this regime is outside the scope of this work, we have shown in Section~\ref{Sec:semiclassics} that the $\mathbb{Z}_2$ exchange symmetry is unbroken if we put QCD(BF) on $\mathbb{R}^2\times T^2$ with the nontrivial 't~Hooft flux as long as $NL\Lambda\ll 1$. 
We also show that the phase diagram obtained in this semiclassical regime satisfies all the anomaly matching/global inconsistency conditions, and thus the small-$L$ and large-$L$ regimes can be adiabatically connected without any phase transitions. 
Therefore, the unbroken exchange symmetry, indicated by the semiclassical analysis, positively supports the validity of the $\mathcal{N} = 1$ SYM/QCD(BF) orbifold equivalence under the assumption of the adiabatic continuity. 

Let us comment on other supports for the unbroken $\mathbb{Z}_2$ exchange symmetry in the previous literature. 
In Ref.~\cite{Shifman:2008ja}, QCD(BF) is put on the small $\mathbb{R}^3\times S^1$ with the double-trace deformation to maintain the center symmetry (see also Refs.~\cite{Unsal:2007jx, Unsal:2007vu, Unsal:2008ch, Poppitz:2012sw, Dunne:2016nmc}). 
In this setup, the confinement is caused by the Coulomb gas of monopole-instantons, and this confinement vacua turn out to break only the chiral symmetry, so the $\mathbb{Z}_2$ exchange symmetry is unbroken. 
It is notable that our semiclassical description of confinement is based on the center vortices and thus the different semiclassical confinement mechanisms yield the same answer. 
Furthermore, one can directly calculate the four-dimensional phase diagram when the fermion is sufficiently heavy ($m\gg \Lambda$), and it is shown in Ref.~\cite{Karasik:2019bxn} that the $\mathbb{Z}_2$ exchange symmetry is again unbroken when $m\gg \Lambda$. 
All these results coherently suggest the unbroken $\mathbb{Z}_2$ exchange symmetry and the validity of the large-$N$ orbifold equivalence, while it is logically possible that all these regions could be separated by phase transitions from the four-dimensional QCD(BF) with the small/intermediate fermion mass. 
All in all, it is plausible to conclude that the QCD(BF) vacuum respects the $\mathbb{Z}_2$ exchange symmetry. 

It would be interesting to directly compare $\mathcal{N} = 1$ super Yang-Mills theory with the exchange-symmetric sector ($g_1 = g_2$ and $\theta_1 = \theta_2 = \theta_d$) of the QCD(BF).
Since the action of the parent theory is reduced to twice the action of the daughter theory, $S_{SYM} \rightarrow 2S$, we note that the correspondence between parameters of the parent and daughter theories is given by 
$\theta_p = 2 \theta_d$ \cite{Kovtun:2005kh}. 
Let us now look into the vacuum structure predicted by the semiclassical descriptions:

\begin{itemize}

    \item Daughter theory: $SU(N)_1 \times SU(N)_2$ QCD(BF)\\
There are $N^2$ classical vacua $\{ \ket{k_1,k_2} \}_{k_1,k_2 = 0, \cdots, N-1}$, in which the exchange-symmetric ones are $N$, $\{ \ket{k,k} \}_{k = 0, \cdots, N-1}$.
This shows indeed the $N$-fold degeneracy of vacua for the massless case on the $\theta_1 = \theta_2 ~(= \theta_d)$ line.
For small $m$ and $k_1=k_2$, the vacuum energy becomes
\begin{align}
    \mathcal{E}_{k,k}(\theta_d) = - \frac{8 m K \rme^{-S_\mathrm{I}/N}}{4 \pi} \log \left( \frac{\Lambda_\mathrm{UV}^2}{4K^2 \rme^{-2S_\mathrm{I}/N}} \right) \cos \left( \frac{\theta_d - 2 \pi k}{N} \right) + O(m^2),
\end{align}
up to some constant. Using the $2$-loop definition of the strong scale $\Lambda=\Lambda_{2\text{-loop}}=\mu\left(\frac{16\pi^2}{Ng^2}\right)^{1/3+1/(9N^2)}\rme^{-8\pi^2/(3Ng^2)}$ (see Appendix~\ref{app:2loopRG}), we obtain 
\begin{equation}
    \mathcal{E}_{k,k}(\theta_d)\sim - m \Lambda(\Lambda L)^2 \left(\ln \frac{1}{\Lambda L}\right)^{-1/3N^2}\cos\left(\frac{\theta_d-2\pi k}{N}\right),
    \label{eq:VacuumEenergyQCDBF}
\end{equation}
with the rough estimation $\Lambda_{\mathrm{UV}}\sim 1/L$ and $K\sim 1/L$. 
We note that $\frac{1}{L^2}\left.\frac{\partial \mathcal{E}_{k,k}}{\partial m}\right|_{m=0}$ gives the scalar chiral condensate in the chiral limit at the $|k,k\rangle$ vacuum, and it almost gives the scale-invariant result, $\sim \Lambda^3$, up to the mild logarithmic correction, $(\ln 1/(\Lambda L))^{-1/3N^2}$, in the prefactor.

    \item Parent theory: $\mathcal{N} = 1$ $SU(2N)$ super Yang-Mills theory\\
    The semiclassical description of the $SU(2N)$ $\mathcal{N} = 1$ super Yang-Mills theory is given in \cite{Tanizaki:2022ngt}.
For the massless case, there are degenerate $2N$ vacua $\{ \ket{k} \}_{k = 0, \cdots, 2N - 1}$ with chiral condensate $\braket{k|\operatorname{tr} \lambda \lambda |k} \sim \Lambda^3 \rme^{\frac{\im(\theta_p - 2 \pi k)}{2N}}$, where $\lambda$ is the gluino field and $\theta_p$ is the theta angle of the parent theory.
By introducing mass perturbation $\frac{m}{g^2} (\operatorname{tr} \lambda \lambda + \operatorname{tr} \bar{\lambda}\bar{\lambda})$, the $2N$ vacua split: the $k$-th vacuum has energy density 
\begin{align}
    \mathcal{E}_k(\theta_p) \sim - m \Lambda (\Lambda L)^2 \cos \left( \frac{\theta_p - 2 \pi k}{2N} \right).
    \label{eq:VacuumEnergySYM}
\end{align}
\end{itemize}

Comparing \eqref{eq:VacuumEenergyQCDBF} and \eqref{eq:VacuumEnergySYM}, 
we can speculate the obvious correspondence between QCD(BF) and $\mathcal{N}=1$ SYM with $\theta_d = \theta_p/2$ as
\begin{align}
    \ket{k,k}_{\text{daughter},\theta_p/2}  ~ \longleftrightarrow ~ \ket{2k}_{\text{parent},\theta_p}.
    \label{eq:equiv_state_even}
\end{align}
For odd vacua of the parent theory, we need another daughter theory at $\theta_d = \theta_p/2 + \pi$,
\begin{align}
    \ket{k,k}_{\text{daughter},\theta_p/2 + \pi}  ~ \longleftrightarrow ~ \ket{2k + 1}_{\text{parent},\theta_p}.
    \label{eq:equiv_state_odd}
\end{align}
This two-to-one correspondence of the vacuum is the one that was pointed out in Ref.~\cite{Kovtun:2005kh}.

\subsubsection{Invalidity of the orbifold equivalence on the domain walls}

Let us comment on the potential violation of the orbifold equivalence on the domain walls. 
In the chiral limit, the parent theory ($\mathcal{N}=1$ $SU(2N)$ SYM) has $2N$ chiral broken vacua, while the daughter theory ($SU(N)\times SU(N)$ QCD(BF)) has only $N$ vacua, and the possible patterns of the domain walls are totally different between these two theories. 
This causes doubt on the validity of the large-$N$ orbifold equivalence itself~\cite{Armoni:2005wta}. 
It is argued in~\cite{Kovtun:2005kh} that this apparent mismatch can be resolved by the multi-valued mapping of the $\theta$ parameter between the parent and daughter theories. 
In the modern perspective, however, the domain walls must support nontrivial dynamics to cancel the anomaly inflow from the bulk, and it seems that domain walls in the parent and daughter theories have different dynamics. 
Here, we would like to settle this subtlety and claim the following: Even when the bulk physics satisfies the orbifold equivalence, the domain-wall physics does not have to. 

Let us discuss how this can be consistent with the standard proof of the large-$N$ orbifold equivalence. 
To see it, we need to consider how we create the domain wall. 
In order to create the domain wall, we usually introduce the spacetime-dependent couplings so that the domain-wall configuration is energetically favored, and then we take the infinite-volume limit. By creating the domain wall, the system gains the energy $\Delta \varepsilon\times V'$, where $\Delta \varepsilon$ is the difference of the energy density between the adiabatically-continued state and the actual ground state and $V'$ is the volume of the sub-region. 
But, the creation of the domain wall costs the energy $T\times A$, where $T$ is the domain wall tension and $A$ is the area of the wall. 
In total, the gain of the energy is 
\begin{equation}
    \Delta E=-\Delta \varepsilon V' +T A. 
    \label{eq:energyGain}
\end{equation}
At finite $N$, this is the whole story, but the subtlety comes in if we also consider the large-$N$ limit. 
We note that 
\begin{equation}
    \Delta \varepsilon\sim O(\Lambda^4), \quad T\sim O(N\Lambda^3),
\end{equation}
and thus the infinite-volume limit and the large-$N$ limit do not commute in this setup. 
When we take the infinite-volume limit first, the first term of \eqref{eq:energyGain} wins and thus the domain-wall state is preferred. 
However, when we take the large-$N$ limit first, the second term of \eqref{eq:energyGain} overcomes and thus the domain wall is not created. 

In the proof of the large-$N$ orbifold equivalence, we always take the large-$N$ limit first and then take the infinite-volume limit. 
In this order of limits, the domain-wall states are never generated, and thus the domain-wall physics cannot be compared in the (at least conventional) large-$N$ orbifold equivalence. 
We believe that this resolves the confusion in the previous literature.

\section{Conclusion}
\label{sec:conclusion}

In this paper, we have examined the phase diagram or vacuum structure of the QCD(BF) using the recently proposed semiclassical center-vortex depiction of the confining vacuum, and this is enabled through anomaly-preserving $T^2$ compactification.

In Section~\ref{sec:global_inconsistency}, we improved constraints on the phase diagram by revisiting the 't~Hooft anomalies and global inconsistencies, specifically concentrating on the $\mathbb{Z}_N$ quotient of the $U(1)_V$ symmetry.
We uncovered a global inconsistency for the $2\pi$ shift of $\theta_- = \theta_1 - \theta_2$, which cannot be detected solely from the $\mathbb{Z}_N^{[1]}$ background.
Subsequently, in Section~\ref{Sec:semiclassics}, we applied the semiclassical description via the anomaly-preserving $T^2$ compactification.
The dilute gas picture of center vortices in the 2d effective theory indicates the presence of $N^2$ semiclassical vacua, with a dynamical fermion present in each semiclassical vacuum. Fluctuations of the dynamical fermion split these $N^2$-fold degenerate semiclassical vacua, and we obtain the QCD(BF) phase diagram as shown at the end of Section~\ref{Sec:semiclassics} for various parameter setups. 
We would like to emphasize that our computation is the rigorous one within the validity of the semiclassical analysis $N\Lambda_{1,2}L\ll1$, and our result covers from the chiral limit to the heavy fermion cases. 
The obtained phase diagrams are consistent with the conjectured ones given in Fig.~\ref{fig:conjectured_phase_boundary}.

In particular, we found that the $\mathbb{Z}_2$ exchange symmetry (or the quiver-permutation symmetry) is not spontaneously broken on the exchange-symmetric line ($g_1 = g_2$ and $\theta_1 = \theta_2$).
According to the necessary and sufficient condition for the nonperturbative equivalence~\cite{Kovtun:2004bz}, our result supports the validity of the $\mathcal{N} = 1$ SYM/QCD(BF) orbifold equivalence under the assumption of the adiabatic continuity.
Incidentally, the comparison of the symmetric sector of QCD(BF) with $\mathcal{N}=1$ SYM theory exhibits the two-to-one correspondence of the vacua in our semiclassically-calculable regime, as suggested in \cite{Kovtun:2005kh}.

We also made some comments on the domain-wall physics in Section~\ref{sec:discussion}. 
Within our $2$d effective theory, all the domain walls connecting different semiclassical vacua become non-dynamical, and we need to insert external Wilson line operators to introduce it. 
We discussed its origin and we found that there are different reasons behind these phenomena depending on which domain wall is considered. 
When the domain wall connects the states with different $1$-form symmetry charges, the universe selection rule requires the introduction of external loop operators. Therefore, the non-dynamical features of those walls are the consequence of the 't~Hooft anomaly. 
For other domain walls connecting the states with the same $1$-form symmetry charges, such a selection rule does not exist and there should exist dynamical domain walls. 
We argued that the dynamical creation of those walls requires the non-zero Kaluza-Klein modes of the bifundamental fermion $\Psi$, and those walls are at least as heavy as $1/NL$. We need to go beyond the zero-mode approximation to study their dynamics in our $T^2$-compactified setup.

The domain wall physics has caused some confusion on the nonperturbative validity of the orbifold equivalence in the previous literature~\cite{Armoni:2005wta, Kovtun:2005kh}. 
Although this is an inaccessible range of our semiclassical computations, we gave the heuristic discussion to resolve this subtlety also in Section~\ref{sec:discussion}. 
We claimed that the domain wall physics is outside the scope of the conventional large-$N$ equivalence whether or not the equivalence holds for the bulk physics. 
It is shown that the domain wall states are never created as long as we take the large-$N$ limit before the infinite-volume limit, and thus the domain wall states cannot be compared between the parent and daughter theories. 
Thus, the potential mismatch of the domain-wall physics does not imply the failure of the nonperturbative large-$N$ equivalence.

There are several interesting future outlooks. One could investigate different rank QCD(BF), specifically, $SU(N_1) \times SU(N_2)$ where $N_1 \neq N_2$. 
It is proposed that this theory enjoys the non-supersymmetric duality cascades~\cite{Karasik:2019bxn}, and it would be interesting to explore it within the semiclassical framework. 
In order to treat QCD(BF) with different ranks, we would need $T^2$ compactification with $U(1)_V$ magnetic flux as in fundamental QCD~\cite{Tanizaki:2022ngt}.
It would be valuable to keep exploring how the semiclassical method works in various gauge theories.

\acknowledgments

The authors thank Mithat \"{U}nsal for useful discussion. 
The work of Y. T. was supported by Japan Society for the Promotion of Science (JSPS) KAKENHI Grant numbers, 22H01218, and by Center for Gravitational Physics and Quantum Information (CGPQI) at Yukawa Institute for Theoretical Physics.
Y.~H. was supported by JSPS Research Fellowship for Young Scientists Grant No.~23KJ1161

\appendix

\section{Two-loop definition of strong scales}
\label{app:2loopRG}

Here, we review the $2$-loop definition of the strong scales for QCD(BF). 
According to Ref.~\cite{Jones:1981we}, the two-loop perturbative renormalization group (RG) equation is given by 
\begin{equation}
    \frac{\diff }{\diff \ln \mu}\alpha_1=-6N \alpha_1^2+(-14N-2)\alpha_1^3+(2N^2-2)\alpha_1^2 \alpha_2, 
\end{equation}
with $\alpha_1=\frac{g_1^2}{16\pi^2}$ and $\alpha_2=\frac{g_2^2}{16\pi^2}$. 
The RG equation for $\alpha_2$ is given by replacement $\alpha_1\leftrightarrow \alpha_2$. 

For simplicitly, let us restrict our attention to the exchange-symmetric case, $\alpha_1=\alpha_2=:\alpha$,  and then the RG equation beceomse 
\begin{equation}
    \frac{\diff \alpha}{\diff \ln \mu}=-6N \alpha^2-(12 N^2+4)\alpha^3, 
\end{equation}
Solving this equation, we obtain the $2$-loop definition of the strong scale as 
\begin{equation}
    \Lambda=\mu \left(\frac{16\pi^2}{Ng^2(\mu)}\right)^{\frac{1}{3}+\frac{1}{9N^2}}\exp\left({-\frac{8\pi^2}{3Ng^2(\mu)}}\right).
\end{equation}
We note that, thanks to the perturbative large-$N$ orbifold equivalence, the leading $N$ behaviors of the beta functions are the same with those of $\mathcal{N}=1$ SYM.

\bibliographystyle{JHEP}
\bibliography{./bifundamental.bib,./QFT.bib}

\end{document}